\documentclass[11pt,prd,onecolumn,amsmath,amssymb,aps,floats,floatfix,nofootinbib]{revtex4-2}
\usepackage[colorlinks=true,urlcolor=blue,anchorcolor=blue,citecolor=blue,filecolor=blue,linkcolor=blue,menucolor=blue,linktocpage=true]{hyperref} % should be commented out if the tex file will be compiled with latex in arXiv!!! (pdflatex is fine)

\usepackage[inline]{enumitem}
\usepackage[multidot]{grffile}  % allow the name of figures to include dots
\usepackage{dcolumn}
\usepackage{bm}
\usepackage{amsmath}
\usepackage{amsfonts}
\usepackage{amssymb}
\usepackage{appendix}
\usepackage{color}
\usepackage{comment}
\usepackage[table]{xcolor}
\usepackage{float}
\usepackage{latexsym}
\usepackage{slashed} % slash mark
\usepackage{pstricks}
\usepackage{indentfirst}
\usepackage{mathrsfs}
\usepackage{multirow}
\usepackage{epsfig,psfrag}
\usepackage{graphicx}
\usepackage{enumitem}
\usepackage{geometry,amssymb,yfonts}
\usepackage{yhmath}
\usepackage{anysize}
\usepackage{subfigure}
\usepackage{mathtools}
\usepackage{setspace} % spacing
\usepackage[utf8]{inputenc} % accept utf-8 input encoding
\usepackage[scientific-notation=true]{siunitx} % comprehensive units
\usepackage{url}
\usepackage{makecell}
\usepackage[normalem]{ulem}

\graphicspath{{fig/}}

\allowdisplaybreaks % allow eqnarray breaks
\begin{document}

\title{Imprints of an early matter-dominated era arising from dark matter dilution mechanism on cosmic string dynamics and gravitational wave signatures}

\author{Shi-Qi Ling}
\author{Zhao-Huan Yu}\email[Corresponding author. ]{yuzhaoh5@mail.sysu.edu.cn}
\affiliation{School of Physics, Sun Yat-Sen University, Guangzhou 510275, China}

\begin{abstract}
We investigate the influence of an early matter-dominated era in cosmic history on the dynamics of cosmic strings and the resulting stochastic gravitational waves. Specifically, we examine the case where this era originates from the dark matter dilution mechanism within the framework of the minimal left-right symmetric model. By numerically solving the Boltzmann equations governing the energy densities of the relevant components, we meticulously analyze the modifications to the cosmological scale factor, the number density of cosmic string loops, and the gravitational wave spectrum. Our results reveal that the early matter-dominated era causes a characteristic suppression in the high-frequency regime of the gravitational wave spectrum, providing distinct and testable signatures for future ground-based interferometer experiments.
\end{abstract}

\maketitle
\tableofcontents

\clearpage

\section{Introduction}

Modern cosmology originates from the study of the universe expansion and element formation~\cite{Gamow:1946eb}.
Further observations have given rise to the standard $\Lambda\mathrm{CDM}$ cosmological model, which suggests that after inflation and reheating, the universe successively experiences a radiation-dominated (RD) era, a matter-dominated era (MD), and a dark-energy-dominated era.
Main evidences involve the observations of the cosmic expansion~\cite{Hubble:1929ig}, the big bang nucleosynthesis (BBN)~\cite{Wagoner:1966pv, Fields:2019pfx}, the cosmic microwave background~\cite{Penzias:1965wn, COBE:1992syq, WMAP:2012nax, Planck:2018vyg}, and the accelerating expansion in recent eras~\cite{SupernovaSearchTeam:1998fmf, SupernovaCosmologyProject:1998vns}.

Nevertheless, these observations hardly date back to pre-BBN eras.
Therefore, it is essential to maintain an open perspective regarding the early history of the universe before BBN~\cite{Allahverdi:2020bys}.
Various hypotheses beyond the standard cosmic history have been put forward, such as an early matter-dominated (EMD) era~\cite{Polnarev:1982, McDonald:1989jd, Assadullahi:2009nf, Erickcek:2015jza, Nelson:2018via, Cirelli:2018iax, Chattopadhyay:2022fwa, Banerjee:2024caa}, a kination-dominated era~\cite{Spokoiny:1993kt, Joyce:1996cp, Ferreira:1997hj, Co:2021lkc, Gouttenoire:2021jhk, Ghoshal:2022ruy}, and an intermediate inflationary era~\cite{Silk:1986vc, Lyth:1995ka, Creminelli:2001th, Konstandin:2011dr}.
The verification of such novel hypotheses necessitates observational probes capable of accessing epochs prior to BBN, when the universe was opaque to photons and traditional electromagnetic detection methods are ineffective.
Nonetheless, this challenge can be addressed through observations of gravitational waves (GWs), a new messenger first detected in 2015~\cite{LIGOScientific:2016aoc}.
Unlike electromagnetic radiation, GWs can propagate freely through space, preserving information from the early universe and allowing them to reach present day observatories.

The stochastic gravitational wave background (SGWB) originating from cosmic strings (CSs) provides a compelling avenue of investigating pre-BBN history~\cite{Cui:2017ufi, Cui:2018rwi, Guedes:2018afo, Ramberg:2019dgi, Auclair:2019wcv, Chang:2019mza, Gouttenoire:2019kij, Gouttenoire:2019rtn, Chang:2021afa, Muia:2023wru, Qiu:2023wbs, 
 Ghoshal:2023sfa}.
CSs are one-dimensional topological defects predicted by various beyond the standard model (SM) theories, particularly those involving a spontaneously broken $\mathrm{U}(1)$ symmetry~\cite{Nielsen:1973cs, Kibble:1976sj}.
CS loops are expected to persist as long-lasting sources, emitting GWs from their formation epoch to the present day.
Once formed, the CS network rapidly evolves into a scaling regime~\cite{Bennett:1989ak, Albrecht:1989mk, Bennett:1989yp, Allen:1990tv}, where its correlation length scales proportionally with the Hubble radius.
Thus, the resulting SGWB spectrum encodes a wealth of information about cosmic history.
Furthermore, the SGWB generated by CSs spans an exceptionally broad frequency range, making it detectable across a diverse array of GW experiments.
These include pulsar timing arrays (PTAs) operating at $10^{-9}\text{--}10^{-7}~\si{Hz}$~\cite{Jenet:2009hk, Kramer:2013kea, IPTA:2023ero}, space-borne interferometers sensitive to $10^{-4}\text{--}10^{-1}~\si{Hz}$~\cite{LISA:2017pwj, TianQin:2015yph, TaiJi:2017nsr}, and ground-based interferometers covering $10\text{--}10^3~\si{Hz}$~\cite{LIGOScientific:2014pky, VIRGO:2014yos, Reitze:2019iox, ET:2023epj}.
Thus, future GW experiments have immense potential for unravelling the mysteries of the early universe.

In this work, we explore how an SGWB spectrum from a preexisting CS network is modified by the EMD era that arises within the framework of the dark matter (DM) dilution mechanism~\cite{Doroshkevich:1983, Baltz:2001rq, Asaka:2006ek, Evans:2019jcs, Cosme:2020mck, Chanda:2021tzi, Chaudhuri:2021bmn, Asadi:2021bxp, Nemevsek:2022anh, Nemevsek:2023yjl}.
This mechanism addresses the DM overproduction problem by relying on significant entropy production from the decay of a long-lived particle, referred to as the ``dilutor''.
For the dilution mechanism to be effective, the dilutor must dominate the energy density of the universe for a finite period, thereby introducing an EMD era that interrupts the conventional RD era.
To quantify the impact of the EMD era, we trace the evolution of the total energy density of the universe by solving the relevant Boltzmann equations.
This allows us to derive the resulting SGWB spectrum, which reflects the influence of the EMD era on CS dynamics and GW emissions.

This paper is structured as follows.
In Section~\ref{sec:CS}, we review the formation of CS loops and analyze their number densities during the RD and MD eras, which are essential for assessing the SGWB spectrum induced by CS loops.
In Section~\ref{sec:example}, we investigate an EMD era emerging from the DM dilution mechanism and examine its effects on the evolution of the CS network, the loop number densities, and the resulting SGWB spectrum.
Finally, in Section~\ref{sec:sum}, we summarize our findings.

\section{Gravitational waves originating from cosmic strings}
\label{sec:CS}

In a scalar field theory with a global or gauge $\mathrm{U}(1)$ symmetry, CSs could be formed after the spontaneous breaking of the $\mathrm{U}(1)$ symmetry in the early universe.
They are one-dimensional topological defects concentrating the energy of the scalar field (and the gauge field for the gauged case)~\cite{Hindmarsh:1994re}.
In the Nambu-Goto approximation, CSs are described as infinitely thin objects with tension $\mu$, which is the energy per unit length.
The dimensionless quantity $G\mu$ is commonly used to represent the CS tension.
Considering the dynamics of CSs, these combined factors establish CSs as promising sources of GWs~\cite{Maggiore:2018sht, Auclair:2019wcv}.

\subsection{Gravitational waves from cosmic string loops}

Cosmic strings are generated randomly in the early universe, leading to the formation of the CS network. Long strings with super-horizon lengths intersect to form CS loops, whose relativistic oscillations can effectively emit GWs. Although small structures in long strings can also produce GWs, their contributions are generally negligible, compared to those generated by loops~\cite{Allen:1991bk, Vilenkin:2000jqa, Gouttenoire:2019kij}. In practice, it is sufficient to consider only the stochastic GWs generated by CS loops. 

The GW emission power of CS loops is given by~\cite{Vachaspati:1984gt, Vilenkin:2000jqa}
\begin{equation}
	\label{eqn:P}
	P = \Gamma G\mu^2,
\end{equation}
where $\Gamma$ is estimated to be approximately 50~\cite{Blanco-Pillado:2017oxo}, and $G$ is the Newtonian gravitational constant. The frequencies of GWs emitted by a CS loop of length $l$ are~\cite{Maggiore:2018sht}
\begin{equation}
	\label{eqn:f}
	f_\mathrm{e} = \frac{2n}{l},\quad n \in \mathbb{N}^{+},
\end{equation}
where $n$ denotes the harmonic modes of the loop oscillation.
Thus, the expression for the power can be rewritten as
\begin{equation}
	\label{eqn:pn}
	P = G\mu^2 \sum_n P_n,
\end{equation}
where $P_n$ is the dimensionless emission power in units of $G\mu^2$ for a single mode $n$, which can be estimated by numerical simulations for GW emissions from the RD and MD eras~\cite{Blanco-Pillado:2017oxo}.

Introducing $n_\mathrm{CS}(l,t) \,\mathrm{d}l$ as the number density of CS loops of length $l$ at cosmic time $t$, the energy density of GWs emitted from CS loops per unit time at the emission time $t_\mathrm{e}$ can be expressed as
\begin{equation}\label{drhoGWdt_e}
	\frac{\mathrm{d}\rho_\mathrm{GW}}{\mathrm{d}t}\bigg|_{t_\mathrm{e}} = G\mu^2 \sum_n P_n\int_0^{l_\mathrm{max}} n_\mathrm{CS}(l,t_\mathrm{e})\,\mathrm{d}l.
\end{equation}
Using Eq.~\eqref{eqn:f}, we derive
\begin{equation}
	\label{eqn:rho}
	\frac{\mathrm{d}^2\rho_\mathrm{GW}}{\mathrm{d}t\,\mathrm{d}f}\bigg|_{t_\mathrm{e}} = G\mu^2 \sum_n \frac{2nP_n}{f_e^2}\,n_\mathrm{CS}\left(\frac{2n}{f_e}, t_\mathrm{e}\right).
\end{equation}
Note that this result corresponds to the GW emission time $t_\mathrm{e}$ but GW experiments only receive signals at the present time $t_0$. Hence, the effect of the cosmological redshift must be accounted for, and the present GW frequency is given by $f = a(t_\mathrm{e})f_\mathrm{e}$, while the present GW energy density is $\rho_\mathrm{GW} = \rho_\mathrm{GW}(t_\mathrm{e}) a^4(t_\mathrm{e})$, where $a(t)$ is the scale factor normalized such that $a(t_0)=1$.
Integrating Eq.~\eqref{eqn:rho} over time, we arrive at
\begin{equation}
	\frac{\mathrm{d}\rho_\mathrm{GW}}{\mathrm{d}f} = G\mu^2 \sum_n P_n C_n(f),
\end{equation}
where
\begin{equation}\label{eqn:Cn}
C_n(f) = \frac{2n}{f^2} \int_{t_*}^{t_0}a^5(t)\, n_\mathrm{CS} \left(\frac{2na(t)}{f}, t\right)\mathrm{d}t,
\end{equation} 
with $t_\ast$ denoting the cosmic time when CS loops start to radiate GWs.

The frequency spectrum of the SGWB induced by CS loops is commonly characterized by a dimensionless quantity
\begin{equation}\label{eqn:OGW}
\Omega_\mathrm{GW}(f) \equiv \frac{1}{\rho_\mathrm{c}}\frac{\mathrm{d}\rho_\mathrm{GW}}{\mathrm{d}\ln f}
= \frac{8\pi G^2\mu^2 f}{3H_0^2}\sum_n P_n C_n(f),
\end{equation}
where $\rho_\mathrm{c} = 3H_0^2 /(8\pi G)$ is the present critical density and $H_0 = 100h~\mathrm{km}~\mathrm{s}^{-1}~\mathrm{Mpc}^{-1}$ is the Hubble constant with $h = 0.674\pm 0.005$~\cite{Planck:2018vyg}.
In order to calculate Eq.~\eqref{eqn:OGW}, we need to know the evolution of the scalar factor $a(t)$, which can be computed from
\begin{equation}
	\label{eqn:a}
	\frac{\mathrm{d}a(t)}{\mathrm{d}t} = a(t)H(t),
\end{equation}
where $H(t)$ is the Hubble expansion rate.

In the standard $\Lambda\mathrm{CDM}$ model, the Hubble rate can be expressed as~\cite{Binetruy:2012ze}
\begin{equation}
	H = H_0\sqrt{\Omega_\mathrm{r} \mathcal{G}(z) a^{-4} + \Omega_\mathrm{m} a^{-3} + \Omega_\Lambda},
\end{equation}
where $z = a^{-1} -1$ is the cosmological redshift. $\Omega_\mathrm{r} = 1.68\times (5.38\pm 0.15)\times 10^{-5}$, $\Omega_\mathrm{m} = 0.315\pm 0.007$, and $\Omega_\Lambda = 0.685\pm 0.007$~\cite{ParticleDataGroup:2022pth} are cosmological constants representing the energy fraction of radiation, matter, and dark energy, respectively.
\begin{equation}
	\mathcal{G}(z) = \frac{g_\star(z)g_{\star s}^{4/3}(0)}{g_\star(0)g_{\star s}^{4/3}(z)}
\end{equation}
is a function of the redshift $z$ introduced to account for the changes in relativistic degrees of freedom, where $g_\star(z)$ and $g_{\star s}(z)$ are the effective numbers of relativistic degrees of freedom for the energy and entropy densities, respectively.
Considering the evolution of $g_\star(z)$ and $g_{\star s}(z)$, we can approximate $\mathcal{G}(z)$ as a piecewise function that changes at the epochs of electron-positron annihilation $(z\simeq10^9)$ and QCD phase transition $(z\simeq2\times10^{12})$:~\cite{Binetruy:2012ze}
\begin{equation}\label{eqn:dof}
	\mathcal{G}(z) = \left\{
	\begin{aligned}
		1,\qquad &z < 10^9,\\
		0.83,\qquad &10^9 < z < 2\times 10^{12},\\
		0.39,\qquad &z > 2\times 10^{12}.
	\end{aligned}
	\right.
\end{equation}

However, if an EMD era exists, the evolution of the Hubble rate $H(t)$ would be altered, necessitating modifications to the expressions above.
This situation is discussed in detail in Section~\ref{sec:example}.

\subsection{Number density of cosmic string loops}\label{subsec:nCS}

For calculating the SGWB spectrum induced by CS loops, estimating the loop number density distribution $n_\mathrm{CS}(l,t)$ is crucial.
Two models can be used to evaluate $n_\mathrm{CS}(l,t)$ based on numerical simulations and the scaling nature of the CS network .
The first is the BOS model, introduced by Blanco-Pillado, Olum, and Shlaer~\cite{Blanco-Pillado:2013qja}.
This model uses the horizon distance as the sole kinematic scale to describe the scaling behavior and extrapolates the loop production functions from numerical simulations to derive $n_\mathrm{CS}(l,t)$ in both RD and MD eras.
The second is the LRS model, proposed by Lorenz, Ringeval, and Sakellariadou~\cite{Lorenz:2010sm}.
In this model, the number density distribution of produced loops per unit time is assumed to follow a power law in the scaling regime, and the gravitational backreaction effect is included to reduce loop production below a certain scale.
Compared to the SGWB spectrum produced by the BOS model, the SGWB spectrum predicted by the LRS model has a significantly higher amplitude.
Consequently, the constraint on the CS tension derived from the LIGO-Virgo O3 dataset is more stringent for the LRS model than for the BOS model~\cite{LIGOScientific:2021nrg}.
Additionally, future space-borne interferometers will more easily detect the SGWB predicted by the LRS model~\cite{Luo:2025ewp}.

Nevertheless, these two models rely on the scaling behavior of the CS network, which may be violated by the inclusion of an EMD era.
To account for this nonscaling effect, we instead adopt the velocity-dependent one-scale (VOS) model~\cite{Vilenkin:2000jqa, Martins:1995tg, Martins:1996jp}, which describes the dynamical evolution of the CS network in a generic manner.
This model has also been used to interpret the positive evidence of nHz SGWB reported by PTA experiments~\cite{Bian:2020urb, NANOGrav:2023hvm, Bian:2023dnv}.
Below, we will demonstrate that the predictions of the VOS model align with those of the BOS model when the scaling regime is achieved and appropriate loop production functions are assumed.
Furthermore, the VOS model can be directly applied to cosmic history with an EMD era, which is discussed in Subsection~\ref{subsec:n}.

In the VOS model, the dynamics of the CS network is characterized by two fundamental parameters, the correlation length $L$ and the root-mean-square (RMS) velocity $v$ of string segments.
Thus, the energy density of long strings can be expressed as $\rho = \mu / L^2$, and its evolution in the universe is governed by~\cite{Martins:1995tg}\begin{equation}\label{eqn:VOS}
    \dot{\rho} = - 2H(1+v^2)\rho - \frac{\tilde{c}v}{L} \rho,
\end{equation}
where the friction effect is neglected.
The term $-2H \rho$ accounts for the dilution and stretching of the strings due to cosmic expansion, while $-2 H v^2 \rho$ captures the energy loss caused by velocity redshift.
The term $-\tilde{c}v\rho/L$ represents the energy removed through loop formation, where $\tilde{c}\simeq 0.23$~\cite{Martins:2000cs} is the loop chopping efficiency.
Additionally, the evolution of the RMS velocity $v$ is determined by~\cite{Martins:1995tg}
\begin{equation}
    \label{eqn:VOSv}
    \dot{v} = (1-v^2)\left[\frac{k(v)}{L}-2Hv\right]
\end{equation}
with~\cite{Martins:2000cs}
\begin{equation}
    k(v) = \frac{2\sqrt{2}}{\pi}(1-v^2)(1+2\sqrt{2}v^3)\frac{1-8v^6}{1+8v^6}.
\end{equation}

Introducing the dimensionless quantity $\xi \equiv L/t$, which represents the correlation length $L$ normalized by the cosmic time $t$, the evolution equations above can be rewritten as
\begin{align}
    t\dot{\xi} &= H(1+v^2)t\xi - \xi + \frac{1}{2}\tilde{c}v,\label{eqn:ksi}\\
    t\dot{v} &= (1-v^2)\left[\frac{k(v)}{\xi} - 2H t v\right].\label{eqn:v}
\end{align}
By numerically solving these equations, it is found that the solutions rapidly converge to constant values of $\xi$ and $v$, independent of the initial conditions.
This indicates that the CS network quickly evolves into a linear scaling regime characterized by $L \propto t$.
By setting $\dot{\xi} = \dot{v} = 0$, the scaling solutions for RD and MD eras can be derived as~\cite{Marfatia:2023fvh}
\begin{alignat}{3}
    \xi_\mathrm{r} &= 0.271,\qquad v_\mathrm{r} &= 0.662,\qquad &\text{RD era},
    \label{eqn:scalingregime:RD}
    \\
    \xi_\mathrm{m} &= 0.625,\qquad v_\mathrm{m} &= 0.582,\qquad &\text{MD era}.
    \label{eqn:scalingregime:MD}
\end{alignat}

The increase in the energy density of CS loops $\rho_\circ$ is driven by the energy transfer from long strings, expressed as~\cite{Gouttenoire:2019kij, Auclair:2022ylu, Marfatia:2023fvh}
\begin{equation}
    \label{eqn:rholoops}
    \dot{\rho}_\circ = \frac{\mathcal{F}}{\gamma_v} \frac{\tilde{c}v}{L}\rho = \frac{\mathcal{F} \tilde{c}v\mu}{\gamma_v \xi^3t^3},
\end{equation}
where the Lorentz factor $\gamma_v \equiv (1-v^2)^{-1/2}$ incorporates the energy loss caused by the redshifting of the loop velocity, and $\mathcal{F}$ is a coefficient that characterizes the fraction of slow, large loops responsible for the dominant contribution to GW emissions.
The values of $\mathcal{F}$ are different for the RD and MD eras, denoted as $\mathcal{F}_\mathrm{r}$ and $\mathcal{F}_\mathrm{m}$, respectively. These values can be determined by a comparison with numerical simulations in the scaling regime~\cite{Blanco-Pillado:2011egf, Blanco-Pillado:2013qja}.

The energy in a CS loop of length $l$ is expressed as $\mu l$.
Thus, the loop production function, defined by~\cite{Gouttenoire:2019kij, Marfatia:2023fvh}
\begin{equation}
\mathcal{P}(l,t) = \frac{1}{\mu l}\frac{\mathrm{d}\dot{\rho_\circ}}{\mathrm{d}l},
\end{equation}
quantifies the rate of increase in the number density of CS loops per unit length per unit time.
Because of GW emission, a CS loop of length $l'$ at time $t'$ loses energy $\Gamma G\mu^2 (t-t')$ by time $t$ and shrinks to a length $l = l' - \Gamma G\mu (t-t')$.
Therefore, the loop number density per unit loop length at time $t$ is given by
\begin{equation}\label{nCS_P_int}
    n_\mathrm{CS}(l,t) = \frac{1}{a^3(t)}\int_{t_\mathrm{ini}}^t \mathcal{P}(l',t')\, a^3(t')\,\mathrm{d}t',
\end{equation}
with $l'(t) = l+\Gamma G\mu (t-t')$. $t_\mathrm{ini}$ denotes the initial time of loop production, and the scale factors account for the dilution caused by cosmic expansion.
To determine $n_\mathrm{CS}(l,t)$ and calculate the energy density of the emitted GWs using Eq.~\eqref{drhoGWdt_e}, the loop production function $\mathcal{P}(l,t)$ is required.

During the RD era, numerical simulations~\cite{Blanco-Pillado:2013qja} reveal that most large loops are generated with a length $l$ characterized by the fraction $\alpha_\mathrm{r} = l/L$ of the correlation length $L$.
$\alpha_\mathrm{r}$ satisfies $\alpha_\mathrm{r} \xi_\star \simeq 0.1$, where the quantity with subscript $\star$ is evaluated at the loop production time $t_\star$.
Consequently, the loop production function can be accurately approximated by
\begin{equation}
\mathcal{P}_\mathrm{r}(l,t) = \frac{\mathcal{F}_\mathrm{r}\tilde{c}v}{\gamma_v \alpha_\mathrm{r}\xi^4t^5}\;\delta\left(\alpha_\mathrm{r}\xi-\frac{l}{t}\right), \quad \text{RD era},
\end{equation}
where $\delta(x)$ denotes the Dirac $\delta$ function, which fixes the loop production time $t_\star$ through the relation $\alpha_\mathrm{r}\xi_\star = {l_\star}/{t_\star}$.
Substituting the expression for $\mathcal{P}_\mathrm{r}(l,t)$ into Eq.~\eqref{nCS_P_int} with $a \propto t^{1/2}$ in the RD era, we derive an analytical expression for the loop number density~\cite{Sousa:2013aaa, Marfatia:2023fvh, Auclair:2022ylu}:
\begin{equation}\label{eqn:ncs_ana_r}
    n^\mathrm{r}_\mathrm{CS}(l,t) = \frac{\mathcal{F}_\mathrm{r}\tilde{c}v_\star}{t^4\gamma_{v_\star}\alpha\xi_\star^4}\frac{\Theta(\alpha\xi_\star - l/t)}{\alpha\xi_\star + \alpha\dot{\xi}_\star t_\star + \Gamma G\mu}\left(\frac{t}{t_\star}\right)^{5/2}, \quad \text{RD era},
\end{equation}
where $\Theta(x)$ represents the Heaviside step function.
By accounting for the decrease in loop length due to GW emission, we have $\alpha_\mathrm{r}\xi_\star {t_\star} = {l_\star} = l + \Gamma G \mu(t - t_\star)$, leading to
\begin{equation}
t_\star = \frac{l + \Gamma G\mu t}{\alpha_\mathrm{r}\xi_\star + \Gamma G\mu}.
\end{equation}
In the scaling regime of the RD era, we have $\xi_\star = \xi_\mathrm{r}$, $v_\star = v_\mathrm{r}$, and $\dot{\xi}_\star = 0$.
Adopting
\begin{equation}
\mathcal{F}_\mathrm{r} = 0.1,
\end{equation}
the loop number density formed in an RD era for $\alpha_\mathrm{r} \xi_\star = 0.1 \gg \Gamma G \mu$ reduces to
\begin{equation}\label{eqn:nr}
n_\mathrm{CS}^\mathrm{r}(l,t) \simeq \frac{0.18}{t^{3/2}(l + \Gamma G\mu t)^{5/2}} \, \Theta(0.1 t - l),\quad \text{scaling, RD era,}
\end{equation}
which agrees with the result derived from numerical simulations in Ref.~\cite{Blanco-Pillado:2013qja}.

In contrast, numerical simulations for the MD era suggest that loop production occurs on a wider range of scales compared to that in the RD era and can be effectively described by a power-law distribution of $l^{-1.69}$ with a cutoff~\cite{Blanco-Pillado:2013qja}.
This motivate us to assume the loop production function as
\begin{equation}
\mathcal{P}_\mathrm{m}(l,t) = \frac{\mathcal{F}_\mathrm{m}\tilde{c}v}{\gamma_v(l/t)^{1.69} \xi^3 t^5}\;\Theta\left(\alpha_\mathrm{m}\xi-\frac{l}{t}\right), \quad \text{MD era},
\end{equation}
where the theta function has a cutoff at $l = \alpha_\mathrm{m} L$, with $\alpha_\mathrm{m}$ satisfying $\alpha_\mathrm{m} \xi_\star \simeq 0.18$~\cite{Blanco-Pillado:2013qja}.
Using $a \propto t^{2/3}$ in the MD era, we derive the loop number density
\begin{equation}\label{eq:nCSm:1}
n^\mathrm{m}_\mathrm{CS}(l,t) = \frac{\mathcal{F}_\mathrm{m} \tilde{c}}{t^2} \int_{t_\mathrm{m}}^t \frac{v'}{\gamma_{v'} \xi'^3 l'^{1.69} t'^{1.31}} \;\Theta\left( t' - \frac{l'}{\alpha_\mathrm{m} \xi} \right) \mathrm{d}t',
\end{equation}
where $t_\mathrm{m}$ is the initial time of the MD era.
Under the assumption that the $t'$-dependence of $\xi'$ and $v'$ is negligible and that $l' = l + \Gamma G \mu(t - t') \simeq l$ for $\Gamma G \mu \ll l/(t - t')$, we perform integration by parts to obtain
\begin{eqnarray}
n^\mathrm{m}_\mathrm{CS}(l,t) &\simeq &\frac{\mathcal{F}_\mathrm{m} \tilde{c}}{t^2} \left[ \left. -\frac{v'}{0.31 \gamma_{v'} \xi'^3 l'^{1.69} t'^{0.31} }\right|^t_{t_\mathrm{m}} +\int_{t_\mathrm{m}}^t \frac{v' \delta [ t' - l'/(\alpha_\mathrm{m} \xi) ]\, \mathrm{d}t' }{0.31 \gamma_{v'} \xi'^3 l'^{1.69} t'^{0.31}} \right] \Theta\left( t - \frac{l}{\alpha_\mathrm{m} \xi} \right)
\nonumber\\
&=& \frac{\mathcal{F}_\mathrm{m} \tilde{c}}{0.31 t^2} \left[ -\frac{v}{\gamma_{v} \xi^3 l^{1.69} t^{0.31} } + \frac{v_\star}{\gamma_{v_\star} \xi_\star^3 l_\star^{1.69} t_\star^{0.31}} \right] \Theta\left( t - \frac{l}{\alpha_\mathrm{m} \xi_\star} \right)
\nonumber\\
&& + \frac{\mathcal{F}_\mathrm{m} \tilde{c}}{0.31 t^2} \frac{v(t_\mathrm{m})}{\gamma_{v(t_\mathrm{m})} \xi^3(t_\mathrm{m}) [l + \Gamma G\mu(t - t_\mathrm{m})]^{1.69} t_\mathrm{m}^{0.31} }\; \Theta\left( t_\mathrm{m} - \frac{l + \Gamma G\mu(t - t_\mathrm{m})}{\alpha_\mathrm{m} \xi_\star} \right),
\label{eq:nCSm:2}
\end{eqnarray}
where $t_\star$ is determined by $\alpha_\mathrm{m}\xi_\star {t_\star} = l + \Gamma G \mu(t - t_\star)$.
For $t \gg t_\mathrm{m}$, the last term in the above expression can be omitted, leading to
\begin{equation}
n^\mathrm{m}_\mathrm{CS}(l,t) \simeq \frac{\mathcal{F}_\mathrm{m} \tilde{c}}{0.31 t^2} \left[ -\frac{v}{\gamma_{v} \xi^3 l^{1.69} t^{0.31} } + \frac{v_\star}{\gamma_{v_\star} \xi_\star^3 l_\star^{1.69} t_\star^{0.31}} \right] \Theta\left( t - \frac{l}{\alpha_\mathrm{m} \xi_\star} \right) .
\end{equation}
Moreover, for small $\Gamma G\mu$, we can use the approximations $t_\star \simeq (\alpha_\mathrm{m}\xi_\star)^{-1}(l + \Gamma G\mu)$ and $t' \simeq (l/t')^{-1}(l + \Gamma G\mu)$ to express the loop number density in the MD era as
\begin{equation}\label{eqn:ncs_ana_m}
    n^\mathrm{m}_\mathrm{CS}(l,t) \simeq \frac{\mathcal{F}_\mathrm{m}\tilde{c}}{0.31t^2(l+\Gamma G\mu t)^2} \left[\frac{v_\star(\alpha_\mathrm{m} \xi_\star)^{0.31}}{\gamma_{v_\star}\xi_\star^3} - \frac{v(l/t)^{0.31}}{\gamma_v\xi^3}\right]\Theta\left(\alpha_\mathrm{m}\xi_\star-\frac{l}{t}\right), \quad \text{MD era}.
\end{equation}
In the scaling regime, where $\xi = \xi_\star = \xi_\mathrm{m}$ and $v = v_\star = v_\mathrm{m}$, taking $\alpha_\mathrm{m} \xi_\star = 0.18$ and 
\begin{equation}
\mathcal{F}_\mathrm{m} = 0.316,
\end{equation}
the loop number density formed in the MD era becomes
\begin{equation}\label{eqn:nm}
	n_\mathrm{CS}^\mathrm{m}(l,t) = \frac{0.27-0.45(l/t)^{0.31}}{t^2(l + \Gamma G\mu t)^2}\,\Theta(0.18t-l),\quad \text{scaling, MD era,}
\end{equation}
which is also consistent with the simulation results in Ref.~\cite{Blanco-Pillado:2013qja}.

We now analyze the SGWB spectrum generated by CS loops within the standard $\Lambda\mathrm{CDM}$ cosmological model, utilizing Eqs.~\eqref{eqn:nr} and \eqref{eqn:nm} for the loop number density in the scaling regime.
In the $\Lambda\mathrm{CDM}$ model, the MD era succeeds the RD era at the time of matter-radiation equality, $t_\mathrm{eq} = 51.1\pm 0.8 ~\mathrm{kyr}$~\cite{ParticleDataGroup:2022pth}.
Notably, CS loops formed during the RD era and surviving into the MD era also contribute to the loop number density in the MD era, as given by~\cite{Blanco-Pillado:2013qja}
\begin{equation}
	\label{eqn:nrm}
	n_\mathrm{CS}^\mathrm{rm}(l,t) = \frac{0.18t_\mathrm{eq}^{1/2}}{t^2(l + \Gamma G\mu t)^{5/2}}\,\Theta(0.1t_\mathrm{eq}-l-\Gamma G\mu t),\quad \text{scaling.}
\end{equation}
By inserting the scale factor $a(t)$ and the loop number density $n_\mathrm{CS}(l,t)$ into Eqs.~\eqref{eqn:Cn} and \eqref{eqn:OGW}, the SGWB spectrum can be obtained.

\begin{figure}[!t]
    \centering
    \includegraphics[width=0.5\textwidth]{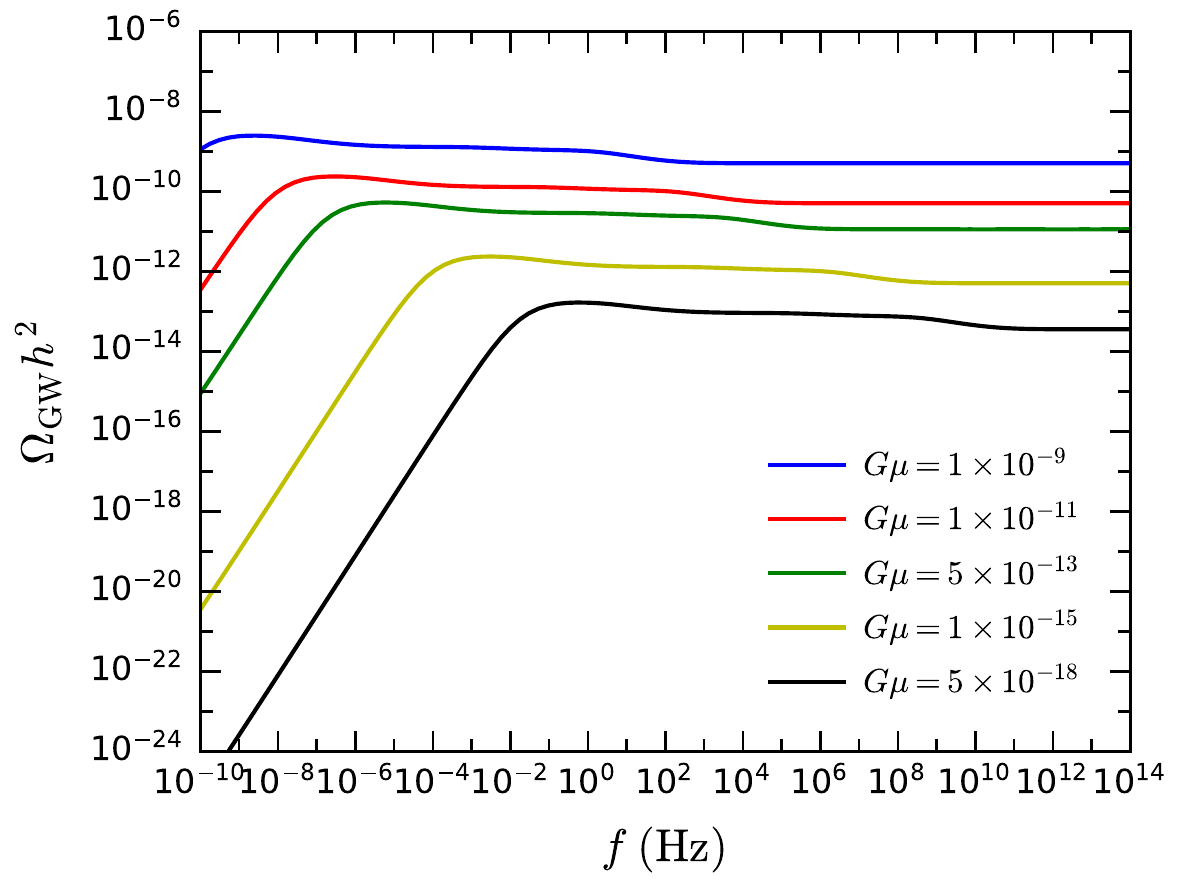}
    \caption{SGWB spectra for various values of $G\mu$ in the standard $\Lambda\mathrm{CDM}$ cosmological model.}
    \label{fig:CSGW_ori}
\end{figure}

In Fig.~\ref{fig:CSGW_ori}, we demonstrate the SGWB spectra in the $\Lambda\mathrm{CDM}$ model for the CS tension parameter $G\mu = 10^{-9}, 10^{-11}, 5\times 10^{-13}, 10^{-15}$, and $5\times 10^{-18}$.
At sufficiently high frequencies, the spectra tend to be flat, primarily due to GW emissions during the RD era~\cite{Blanco-Pillado:2013qja}.
For smaller values of $G\mu$, the total GW emission power is reduced, allowing CS loops to survive longer.
At a specific cosmic time, the average length of loops would be smaller, leading to GWs emitted at higher frequencies.
As a result, the SGWB spectrum shifts downward in amplitude and rightward in frequency as $G\mu$ decreases~\cite{Blanco-Pillado:2017oxo}.

In the following section, we will examine the effects of an additional EMD era inserted into the RD era, focusing on its impact on the scale factor, loop number density, and SGWB spectrum.
This modification may violate the scaling behavior, necessitating the use of the more general expressions for the loop number density, as in Eqs.~\eqref{eqn:ncs_ana_r} and \eqref{eqn:ncs_ana_m}.

\section{Gravitational wave spectrum influenced by an early matter-dominated era}
\label{sec:example}

In this section, we discuss an EMD era motivated by the DM dilution mechanism and study its impact on the SGWB spectrum generated by cosmic strings.

\subsection{Early matter-dominated era}\label{subsec:EMD}

\begin{figure}[!t]
    \centering
    \includegraphics[width=.8\textwidth]{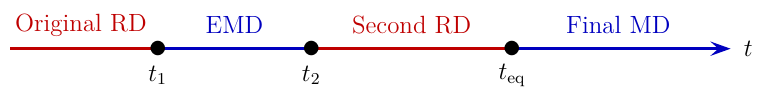}
    \caption{Sketch of the cosmic timeline featuring an EMD era. In chronological order, the universe evolves through the original RD, EMD, second RD, and final MD eras.}
    \label{fig:timeaxis}
\end{figure}

An EMD era is an MD period spanning from cosmic time $t_1$ to $t_2$, embedded within the conventional RD era.
Thus, there exists an initial RD era before $t_1$, and a subsequent RD era after $t_2$.
The second RD era ends at the time of matter-radiation equality $t_\mathrm{eq}$, after which the universe transitions into the final MD era, corresponding to the traditional cosmological evolution.
This timeline is depicted in Fig.~\ref{fig:timeaxis}.

We propose the following assumption for the origin of such an EMD era.
A massive, long-lived particle $Y$ decouples from thermal equilibrium and transitions into nonrelativistic matter, dominating the universe as the temperature decreases and initiating an EMD era at time $t_1$.
The slow decays of $Y$ into SM particles release entropy into the plasma.
In the context of the DM dilution mechanism, this entropy injection dilutes the DM abundance to the observed value.
Once the majority of $Y$ particles have decayed, the EMD era concludes at time $t_2$.

Now, we discuss the evolution of the related energy densities.
The radiation and matter energy densities, $\rho_\mathrm{r}$ and $\rho_\mathrm{m}$, scale with the scale factor $a$ as~\cite{Kolb:1990vq}
\begin{equation}
	\label{eqn:rhoa}
	\rho_\mathrm{r} \propto a^{-4},\quad
	\rho_\mathrm{m} \propto a^{-3}.
\end{equation}
The scale factor evolves as $a \propto t^{1/2}$ in the RD era and $a \propto t^{2/3}$ in the MD era.
Starting from an initial temperature $T_\mathrm{ini}$ in the original RD era, the $Y$ particles constitute a matter component with energy density $\rho_\mathrm{Y}$, while the plasma of SM particles has energy density $\rho_\mathrm{SM}$.
Consequently, the energy densities evolve as
\begin{equation}\label{eqn:rhotr}
\rho_\mathrm{SM} \propto t^{-2},\quad
\rho_\mathrm{Y} \propto t^{-3/2},
\end{equation}
during the original RD era, and 
\begin{equation}\label{eqn:rhotm}
\rho_\mathrm{SM} \propto t^{-8/3},\quad
\rho_\mathrm{Y} \propto t^{-2},
\end{equation}
during the EMD era.
Notably, the energy density of $Y$ decreases more slowly than that of SM particles during the original RD era, leading to the onset of the EMD era.

\subsection{Dark matter dilution mechanism}

In this subsection, we discuss the DM dilution mechanism as the origin of the EMD era.
As severe constraints from DM detection experiments have reduced interest in weakly interacting massive particles with masses around the electroweak scale as DM candidates, a lighter DM candidate $X$ has gained prominence due to its ability to evade direct detection bounds.
Nevertheless, the thermal production of $X$ particles with low annihilation cross sections typically results in
an overproduction problem~\cite{Ellis:1982yb, Ellis:1983ew}.
Within the DM dilution mechanism~\cite{Baltz:2001rq, Asaka:2006ek, Evans:2019jcs, Cosme:2020mck, Chanda:2021tzi, Asadi:2021bxp, Nemevsek:2022anh}, the overproduction of DM particles is mitigated by entropy injection from the decays of a dilutor particle $Y$, which dominates the universe for a period, thereby inducing an EMD era.

Following Ref.~\cite{Nemevsek:2023yjl}, we discuss the DM dilution mechanism within the minimal left-right symmetric model (LRSM) as a concrete and illustrative example.
In this model, the lightest right-handed neutrino $N_1$ serves as the light DM candidate $X$, and the next-to-lightest right-handed neutrino $N_2$ or a neutral Higgs boson $\Delta$ acts as the dilutor $Y$.
As a Majorana fermion, the DM candidate $X$ has two degrees of freedom and its number density in thermal equilibrium and in the relativistic limit is given by~\cite{Kolb:1990vq}
\begin{equation}
	n_X = \frac{3\zeta(3)}{2\pi^2}\, T^3.
\end{equation}
where $T$ denotes the temperature and $\zeta(x)$ represents the Riemann zeta function.
For sufficiently weak interactions, $X$ particles freeze out from the plasma at a temperature $T_\mathrm{FO}\gg m_X$, leaving a number density in the comoving volume of
\begin{equation}
	Y_X \equiv \frac{n_X}{s} = \frac{135\zeta(3)}{4\pi^4g_{\star s}(T_\mathrm{FO})},
\end{equation}
where $s = 2\pi^2 g_{\star s}(T) T^3/45$ is the entropy density.

If $X$ particles evolve without interference, $Y_X$ remains conserved, leading to a present-day relic energy density fraction of~\cite{Nemevsek:2023yjl}
\begin{equation}
	\Omega^0_X = \frac{m_XY_Xs_0}{\rho_\mathrm{c}}\simeq 2.6\, \frac{m_X}{\mathrm{keV}} \frac{100}{g_{\star s}(T_\mathrm{FO})},
\end{equation}
where $s_0 = 2891.2~\mathrm{cm}^{-3}$ and $\rho_\mathrm{c} = 1.05\times 10^{-5} h^2~\mathrm{GeV}/\mathrm{cm}^3$~\cite{ParticleDataGroup:2022pth} are the present time entropy density and the critical energy density, respectively.
Thus, for $m_X \lesssim \si{keV}$, $\Omega^0_X$ exceeds the observed cold DM relic density $\Omega_\mathrm{CDM} = 0.265$~\cite{ParticleDataGroup:2022pth} by at least one order of magnitude, causing an overproduction problem.

Introducing a long-lived dilutor $Y$ with a mass $m_Y$ much larger than $m_X$ can effectively address the problem of $X$ particle overproduction.
First, during the RD era, both $Y$ and $X$ particles decouple relativistically at a similar temperature, resulting in comparable yields, $Y_Y \simeq Y_X$.
Second, because $m_Y\gg m_X$, $Y$ particles become nonrelativistic at a relatively high temperature, while $X$ particles remain relativistic.
Consequently, $Y$ particles quickly dominate the energy density of the universe, initiating an EMD era.
Finally, when the lifetime of $Y$ particles comes to an ends, they decay into SM particles and $X$ particles, injecting entropy and consequently diluting the energy density $\rho_X$ of $X$ particles.

This situation can be analyzed using the Boltzmann equations~\cite{Nemevsek:2022anh}
\begin{align}\label{eqn:Boltzmann1}
	\frac{\mathrm{d}\rho_Y}{\mathrm{d}t} + 3H\rho_Y &= -\Gamma_Y\rho_Y,\\
	\label{eqn:Boltzmann2}\frac{\mathrm{d}\rho_X}{\mathrm{d}t} + 4H\rho_X &= y B_X\Gamma_Y\rho_Y,\\\label{eqn:Boltzmann3}\frac{\mathrm{d}\rho_\mathrm{SM}}{\mathrm{d}t} + 4H\rho_\mathrm{SM} &= (1-y B_X)\Gamma_Y\rho_Y,	
\end{align}
where $\Gamma_Y$ is the decay width of the dilutor $Y$, $y$ represents the energy fraction carried away by $X$ particles among all the decay products of $Y$, and $B_X$ is the branching ratio of the  decay channel $Y \rightarrow X $.
Assuming that $X$ particles remain ultra-relativistic throughout the aforementioned process, and using
\begin{equation}
	\label{eqn:H}
	H = \sqrt{\frac{8\pi}{3M_\mathrm{Pl}^2}(\rho_X + \rho_Y + \rho_{\mathrm{SM}})},
\end{equation}
we can solve the Boltzmann equations. 

In the minimal LRSM with the $\mathrm{SU}(3)_\mathrm{C} \times \mathrm{SU}(2)_\mathrm{L} \times \mathrm{SU}(2)_\mathrm{R} \times \mathrm{U}(1)_{B-L}$ gauge symmetry~\cite{Mohapatra:1974gc, Senjanovic:1975rk} considered in Ref.~\cite{Nemevsek:2023yjl}, two scenarios exist for the DM dilution mechanism, where the DM candidate $X$ is the lightest right-handed neutrino $N_1$.
\begin{itemize}
	\item Scenario 1: the dilutor $Y$ is the next-to-lightest right-handed neutrino $N_2$, which undergoes a three-body decay mediated by a right-handed gauge boson $W^\pm_\mathrm{R}$ into two charged leptons $\ell \ell'$ and one $N_1$. The related right-handed charged current interactions are described by the Lagrangian
	\begin{equation}
		\mathcal{L}_1 = \frac{g}{\sqrt{2}}W_\mathrm{R}^{\mu}\left(\sum_{i=1}^2\bar{N_i}\gamma_\mu V_\mathrm{PMNS}^{R\dag} \ell_\mathrm{R} + \bar{u}_\mathrm{R}\gamma_\mu V_\mathrm{CKM}^\mathrm{R} d_\mathrm{R} \right) + \mathrm{H.c.},
	\end{equation}
    where $g$ is the unified $\mathrm{SU}(2)_\mathrm{L}\times \mathrm{SU}(2)_\mathrm{R}$ gauge coupling, and $V_\mathrm{PMNS}^\mathrm{R}$ and $V_\mathrm{CKM}^\mathrm{R}$ are the right-handed Pontecorvo-Maki-Nakagawa-Sakata (PMNS) and Cabibbo-Kobayashi-Maskawa (CKM) matrices that connect the flavor and mass bases.
   The decay channels of $Y$ include $N_2 \rightarrow N_1\ell\ell'$, $N_2 \rightarrow \ell q\bar{q}'$, and $N_2 \rightarrow \ell W$.
	\item Scenario 2: the dilutor $Y$ is the neutral Higgs boson $\Delta$, which is associated with the origin of the Majorana neutrino masses and can decay into two $N_1$ particles. The relevant Yukawa couplings are given by the Lagrangian
    \begin{equation}
        \mathcal{L}_2 = \bar{Q}_\mathrm{L} (Y_q\Phi + \tilde{Y}_q\tilde{\Phi})Q_\mathrm{R} + \bar{L}_\mathrm{L}(Y_l\Phi + \tilde{Y}_l\tilde{\Phi})L_\mathrm{R} + Y_{\Delta_\mathrm{L}}L^\mathrm{T}_\mathrm{L}\mathrm{i}\sigma_2\Delta_L L_\mathrm{L} + Y_{\Delta_\mathrm{R}}L^\mathrm{T}_\mathrm{R}\mathrm{i}\sigma_2\Delta_\mathrm{R} L_\mathrm{R} + \mathrm{H.c.},
    \end{equation}
    where the family indices are suppressed. $\Phi$ is an $\mathrm{SU}(2)_\mathrm{L}\times \mathrm{SU}(2)_\mathrm{R}$ bidoublet scalar, with $\tilde{\Phi} = \mathrm{i}\sigma_2 \Phi^\ast \mathrm{i}\sigma_2$. $\Delta_\mathrm{L}$ and  $\Delta_\mathrm{R}$ are $\mathrm{SU}(2)_\mathrm{L}$ and $\mathrm{SU}(2)_\mathrm{R}$ triplet scalars, respectively. $Q_{\mathrm{L},\mathrm{R}}$ and $L_{\mathrm{L},\mathrm{R}}$ denote left-handed and right-handed quarks and leptons. $Y_q$, $\tilde{Y}_q$, $Y_l$, $\tilde{Y}_l$, $Y_{\Delta_\mathrm{L}}$, and $Y_{\Delta_\mathrm{R}}$ represent the Yukawa coupling matrices. The dilutor $\Delta$ is the Higgs boson arising from the electrically neutral component of $\Delta_\mathrm{R}$.
    Its decay channels involve $\Delta \rightarrow N_1 N_1$, $\Delta \rightarrow W_\mathrm{R}^\ast W_\mathrm{R}^\ast \to \text{4 fermions}$, $\Delta \rightarrow \gamma \gamma$, etc.
\end{itemize}
We will consider both scenarios in the following analysis.

\begin{table}[!t]
\centering
\setlength\tabcolsep{.6em}
\renewcommand{\arraystretch}{1.3}
\caption{Benchmark points in  Scenario 1 where $N_2$ serves as the dilutor $Y$.}\label{tab:para1}
\begin{tabular}{c|c|c|c}
\hline\hline
Scenario 1 &BP1a   &BP1b   &BP1c   \\
\hline
 Common & \multicolumn{3}{c}{$m_{N_2}= 200~\mathrm{GeV}$,~ $y = 0.35$,~ $\tan\beta = 0.5$}\\
\hline
$m_{N_1}$   &$6.5~\mathrm{keV}$ &$10~\mathrm{keV}$  &$30~\mathrm{keV}$  \\
$m_{W_\mathrm{R}}$   &$5\times 10^7~\mathrm{GeV}$    &$6\times 10^7~\mathrm{GeV}$    &$7\times 10^7~\mathrm{GeV}$    \\
$\Gamma_Y$  &$2.22\times 10^{-23}~\mathrm{GeV}$ &$1.07\times 10^{-23}~\mathrm{GeV}$ &$5.77\times 10^{-24}~\mathrm{GeV}$ \\
$B_X$   &$4.41\times 10^{-3}$   &$4.41\times 10^{-3}$   &$4.41\times 10^{-3}$   \\
\hline\hline
\end{tabular}
\end{table}

\begin{table}[t]
    \centering
    \setlength\tabcolsep{.6em}
    \renewcommand{\arraystretch}{1.3}
    \caption{Benchmark points in  Scenario 2 where $\Delta$ serves as the dilutor $Y$.}\label{tab:para2}
    \begin{tabular}{c|c|c|c}
\hline\hline
\text{Scenario 2} &BP2a   &BP2b   &BP2c   \\
\hline
 Common & \multicolumn{3}{c}{$m_{N_1}= 6.5~\mathrm{keV}$,~ $y = 1$,~ $\theta_{\Delta h} = 0$}\\
\hline
$m_\Delta$   &$1~\mathrm{TeV}$  &$10^3~\mathrm{TeV}$    &$10^6~\mathrm{TeV}$ \\
$m_{W_\mathrm{R}}$   &$10^{11}~\mathrm{GeV}$    &$3.16\times 10^{12}~\mathrm{GeV}$    &$10^{14}~\mathrm{GeV}$    \\
$\Gamma_Y$  &$3.51\times 10^{-21}~\mathrm{GeV}$ &$3.51\times 10^{-15}~\mathrm{GeV}$ &$3.51\times 10^{-9}~\mathrm{GeV}$ \\
$B_X$   &$5.11\times 10^{-12}$   &$5.11\times 10^{-18}$   &$2.42\times 10^{-20}$   \\
\hline\hline
\end{tabular}
\end{table}

We solve the Boltzmann equations for the benchmark points (BPs) specified in Tables~\ref{tab:para1} and~\ref{tab:para2} for Scenarios~1 and 2, respectively.
The values of $y$, $m_{N_1}$, $m_{N_2}$, $m_\Delta$, $m_{W_\mathrm{R}}$, the ratio $\tan\beta$ of two vacuum expectation values from the bidoublet, and the mixing angle $\theta_{\Delta h}$ between $\Delta$ and the $125~\si{GeV}$ Higgs boson $h$ are adopted based on the results presented in Ref.~\cite{Nemevsek:2023yjl}.
The dilutor decay width $\Gamma_Y$ and branching ratio $B_X$ are calculated using the formulas provided therein.
All BPs are selected to ensure the correct DM relic abundance through the DM dilution mechanism.
BP1a and BP2a will be the primary focus of the following analysis, while the remaining BPs will provide additional results for comparison.

\begin{figure}[!t]
    \label{fig:rho_t}
    \centering
    \subfigure[BP1a in Scenario 1\label{fig:rhotHN}]{\includegraphics[width=0.48\textwidth]{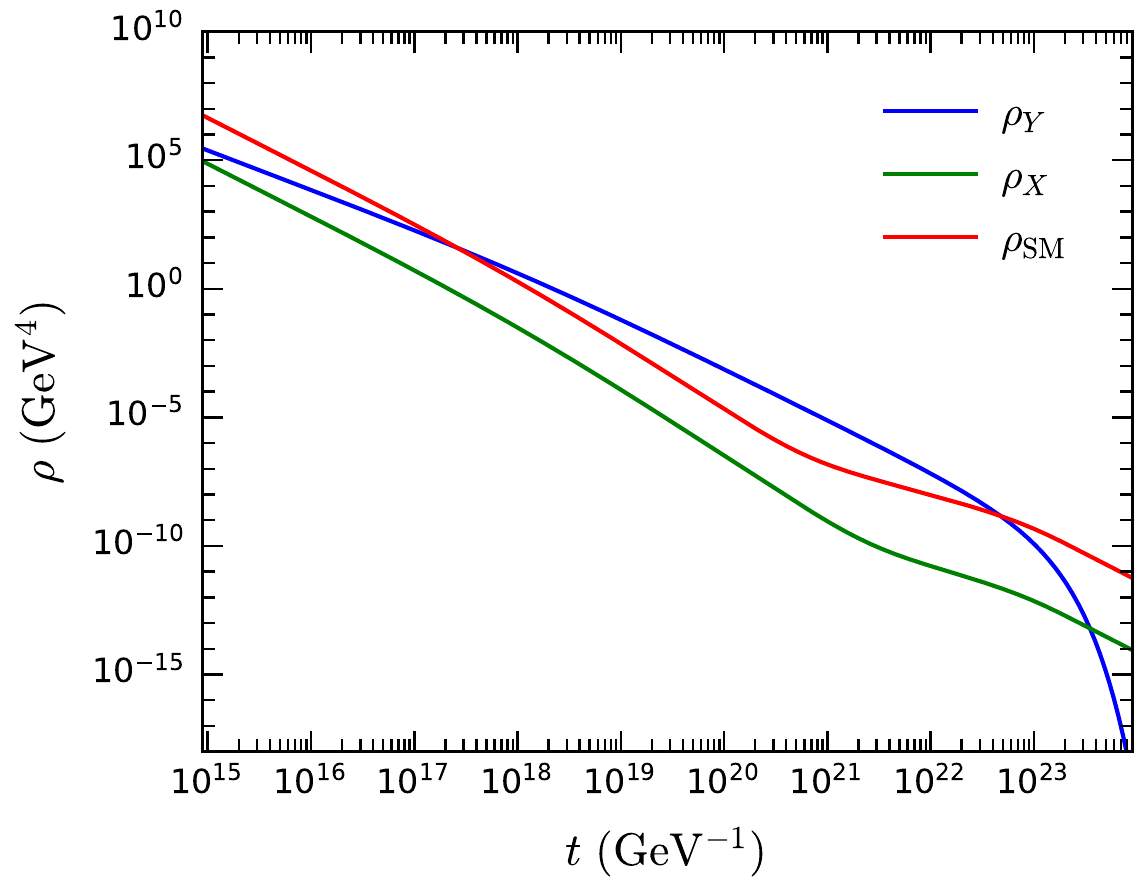}}
    \hspace{.01\textwidth}
    \subfigure[BP2a in Scenario 2\label{fig:rhotMH}]{\includegraphics[width=0.48\textwidth]{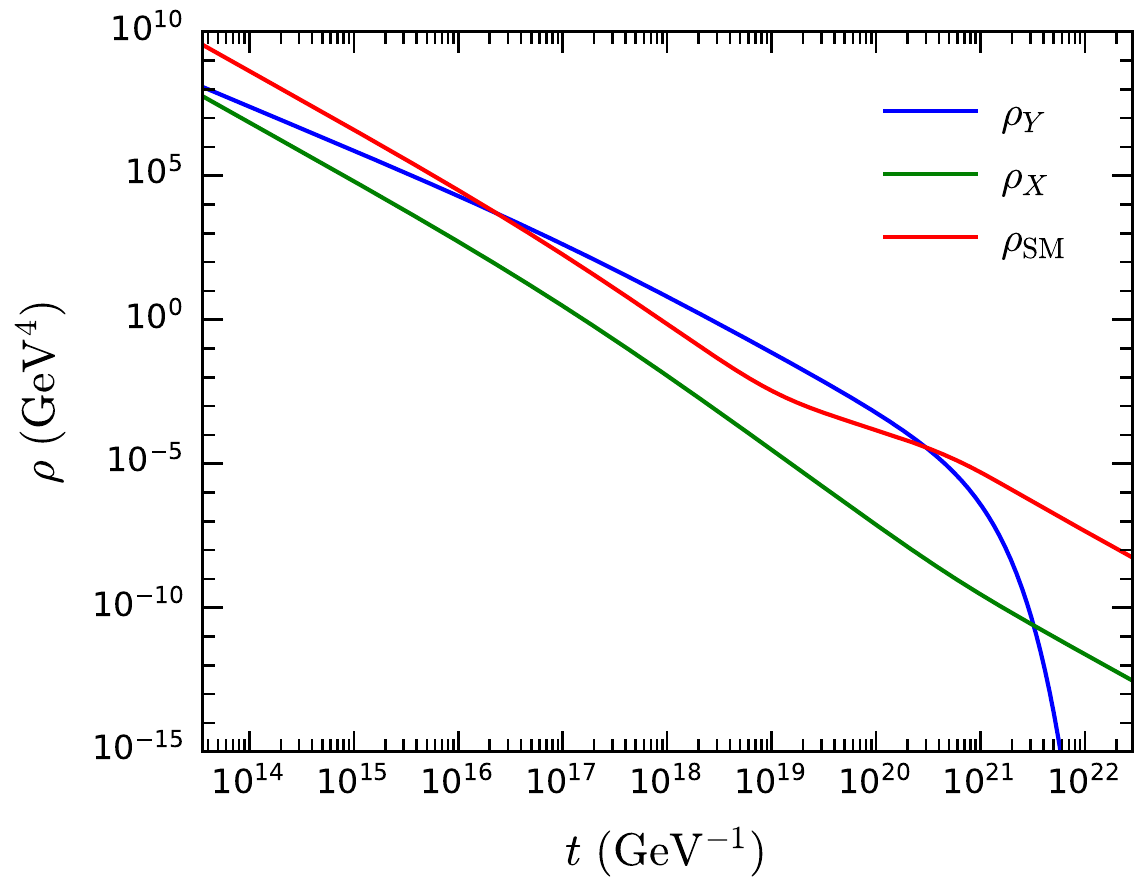}}
    \caption{Evolution of the energy densities $\rho_Y$, $\rho_X$, and $\rho_\mathrm{SM}$ for BP1a in Scenario~1 with the dilutor $N_2$ (a) and BP2a in Scenario~2 with the dilutor $\Delta$ (b).\label{fig:rhot}}
\end{figure}

We adopt the initial temperature as $T_\mathrm{ini} = m_Y/10$, at which $Y$ particles have decoupled with $Y_Y \simeq Y_X$ and are nonrelativistic, while $X$ particles are relativistic.
The initial conditions for the energy densities of $Y$, $X$, and SM particles are
\begin{align}\label{rhoini}
	\rho_{Y}^\mathrm{ini} &= m_Y Y_Y s(T_\mathrm{ini}) \simeq \frac{2\pi^2}{45}\, m_Y Y_Xg_{\star}(T_\mathrm{ini}) T_\mathrm{ini}^3,\\
    \rho_{X}^\mathrm{ini} &=  \frac{7\pi^2}{120}\,T_\mathrm{ini}^4,\\
	\rho_{\mathrm{SM}}^\mathrm{ini} &= \frac{\pi^2}{30}\,g_\star(T_\mathrm{ini}) T_\mathrm{ini}^4.
\end{align}
The obtained energy densities $\rho_Y$, $\rho_X$, and $\rho_\mathrm{SM}$ as functions of cosmic time $t$ are illustrated in Fig.~\ref{fig:rhot} for BP1a in Scenario~1 and BP2a in Scenario~2.

As demonstrated by Figs.~\ref{fig:rhotHN} and \ref{fig:rhotMH}, the initial energy density of the dilutor $Y$ in both BPs is lower than $\rho_\mathrm{SM}$ by one order of magnitude.
Nonetheless, the decrease in $\rho_Y$ is slower than that of $\rho_\mathrm{SM}$ as the universe expands, leading to $\rho_Y = \rho_\mathrm{SM}$ at $t_1 \simeq \num{3e17}~\si{GeV^{-1}}$ and $t_1 \simeq \num{2e16}~\si{GeV^{-1}}$ in BP1a and BP2a, respectively.
This indicates the beginning of the EMD era.
Subsequently, $Y$ particles dominate the universe until their decays become effective, leading to the end of the EMD era at $t_2 \simeq \num{5e22}~\si{GeV^{-1}}$ for BP1a and $t_2 \simeq \num{3e20}~\si{GeV^{-1}}$ for BP2a.
Since $\Gamma_Y$ in BP1a is lower than that in BP2a, as shown in Tables~\ref{tab:para1} and \ref{tab:para2}, the decays of $Y$ particles occur at later times, leading to a longer duration of the EMD era.
The $Y$ decays inject entropy into the plasma, increasing $\rho_\mathrm{SM}$, the plasma temperature, and the entropy density $s$.
Consequently, the $X$ comoving number density $Y_X = n_X/s$ is significantly diluted, ensuring that the relic abundance of $X$ particles is consistent with the observation in both BPs.

\begin{figure}[!t]
	\centering
	\subfigure[BP1a in Scenario 1\label{fig:atHN}]{\includegraphics[width=0.48\textwidth]{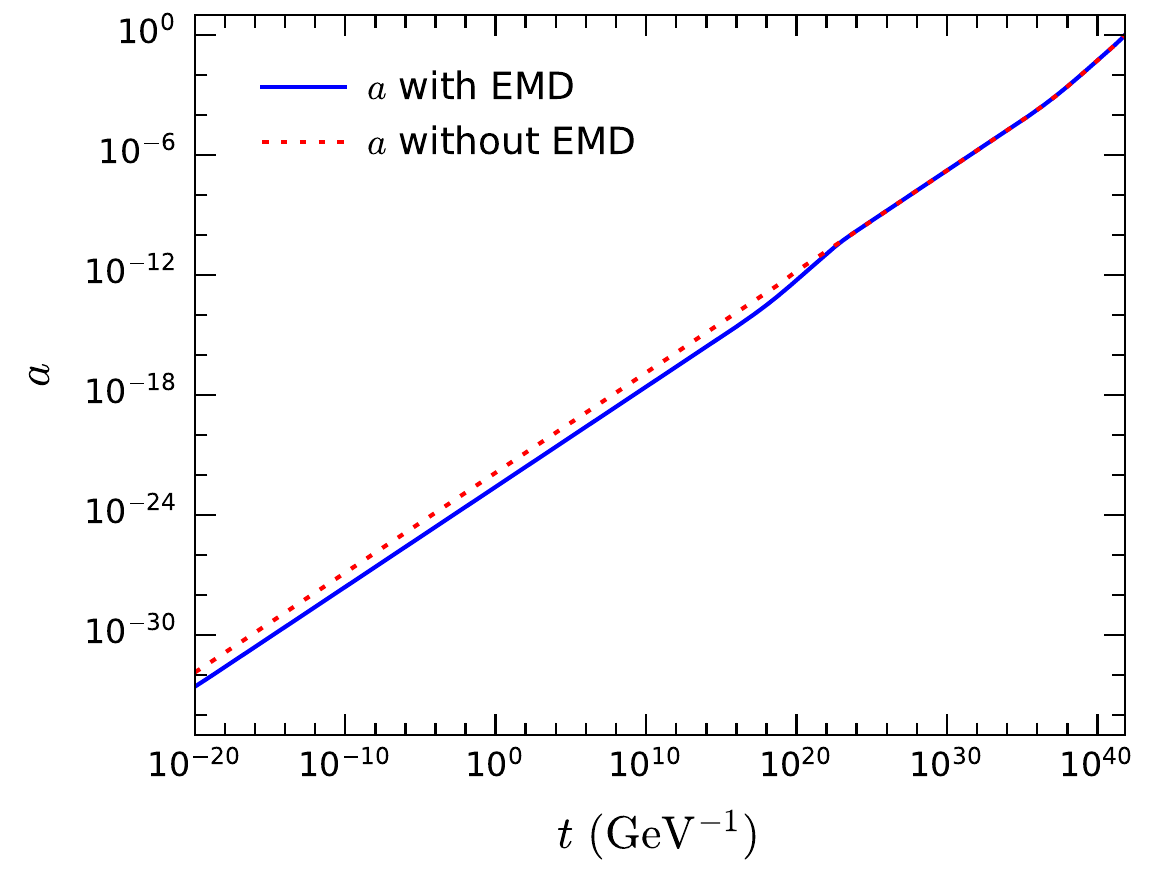}}
	\hspace{.01\textwidth}
	\subfigure[BP2a in Scenario 2\label{fig:atMH}]{\includegraphics[width=0.48\textwidth]{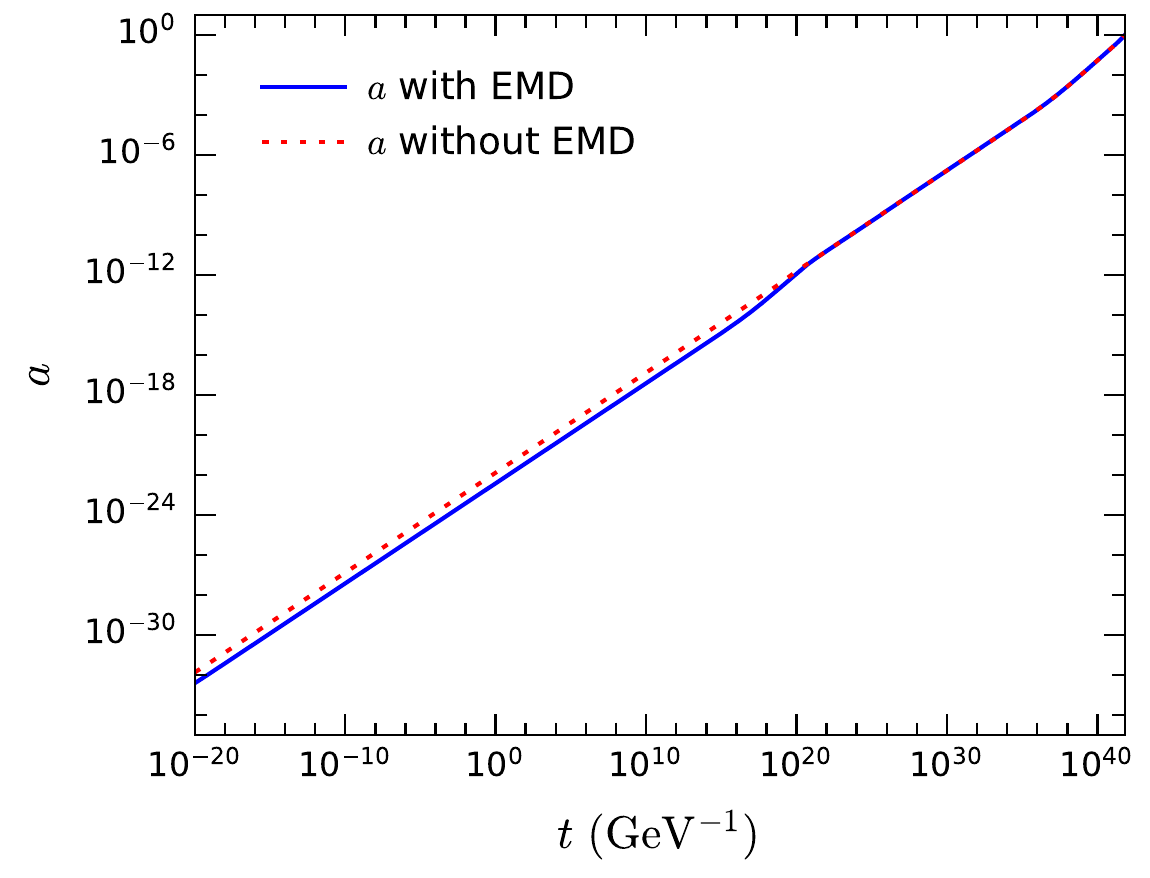}}
	\caption{Evolution of the scale factor $a(t)$ (blue solid lines) for BP1a in Scenario~1 with the dilutor $N_2$ (a) and BP2a in Scenario~2 with the dilutor $\Delta$ (b). For comparison, the result from the standard $\Lambda\mathrm{CDM}$ model, which does not include the EMD era, is represented by red dotted lines.}
 \label{fig:a_h_t}
\end{figure}

By substituting the obtained $\rho_Y$, $\rho_X$, and $\rho_\mathrm{SM}$ into Eq.~\eqref{eqn:H} and solving Eq.~\eqref{eqn:a} with $a(t_0) = 1$, we determine the evolution of the scale factor $a(t)$ in both BPs, as illustrated in Fig.~\ref{fig:a_h_t}.
Compared with the standard $\Lambda\mathrm{CDM}$ cosmological model, the presence of the EMD era reduces the scale factor prior to $t_2$.
This occurs because the scale factor evolves as $a \propto t^{2/3}$ during an MD era, increasing more rapidly with time than $a \propto t^{1/2}$ during an RD era.
Therefore, a smaller $a$ is required at the onset of the EMD era to ensure $a(t_0) = 1$ at the present time $t_0$.

\subsection{Impact on the loop number density of cosmic strings}\label{subsec:n}

\begin{figure}[!t]
	\centering
	\subfigure[$\xi(t)$ for BP1a\label{fig:ksiHN}]{\includegraphics[width=0.48\textwidth]{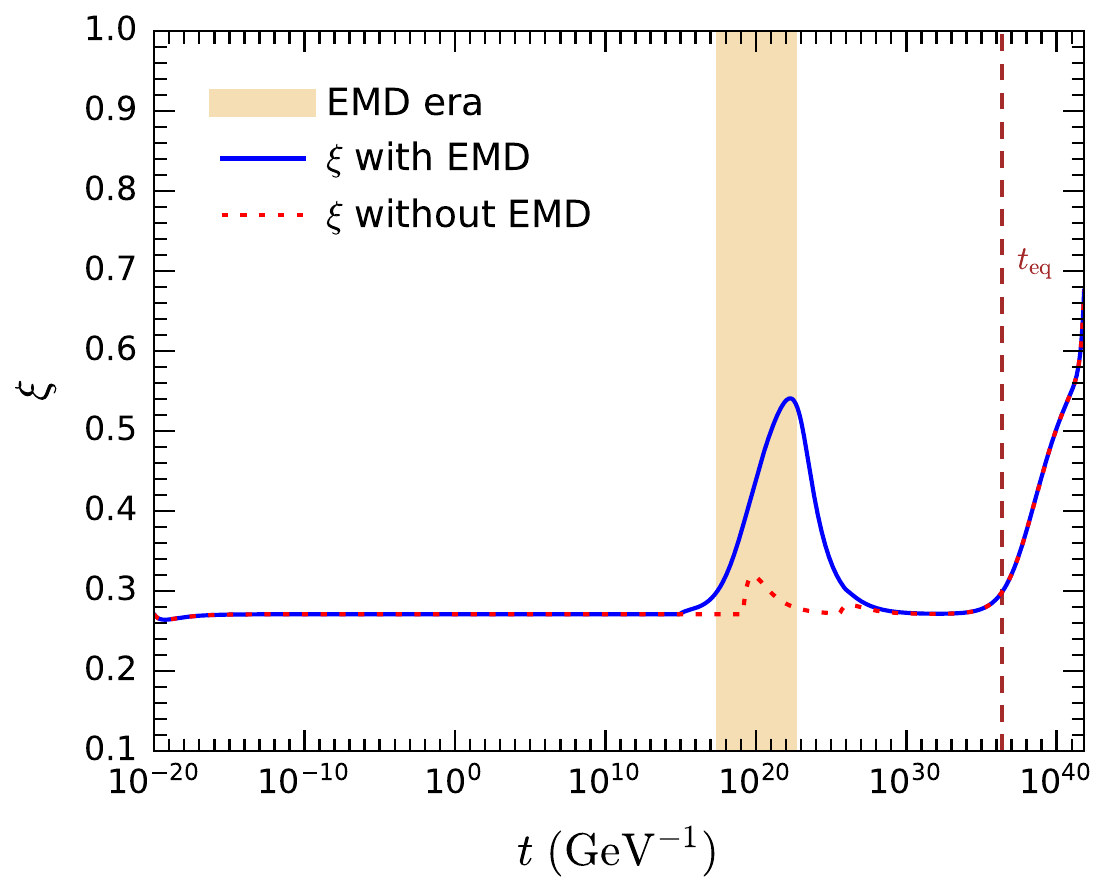}}
	\hspace{.01\textwidth}
	\subfigure[$v(t)$ for BP1a\label{fig:vHN}]{\includegraphics[width=0.48\textwidth]{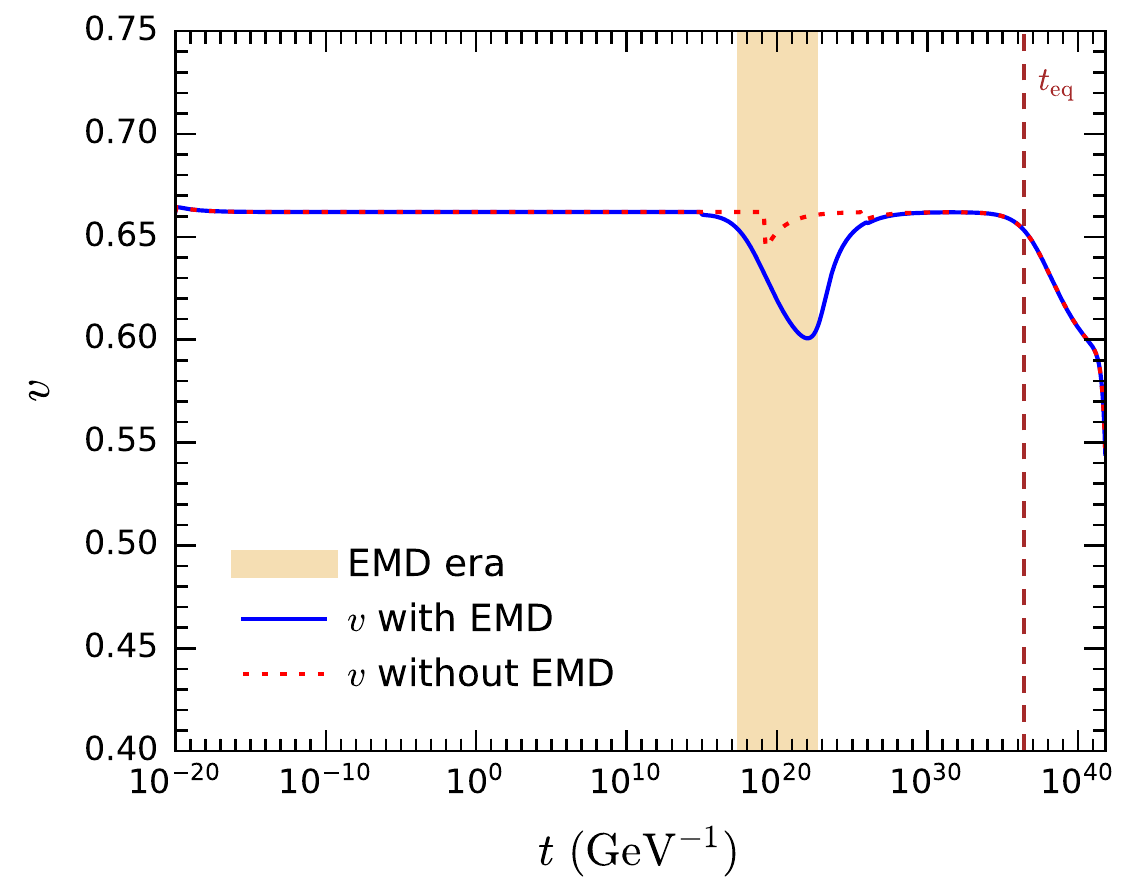}}
    \subfigure[$\xi(t)$ for BP2a\label{fig:ksiMH}]{\includegraphics[width=0.48\textwidth]{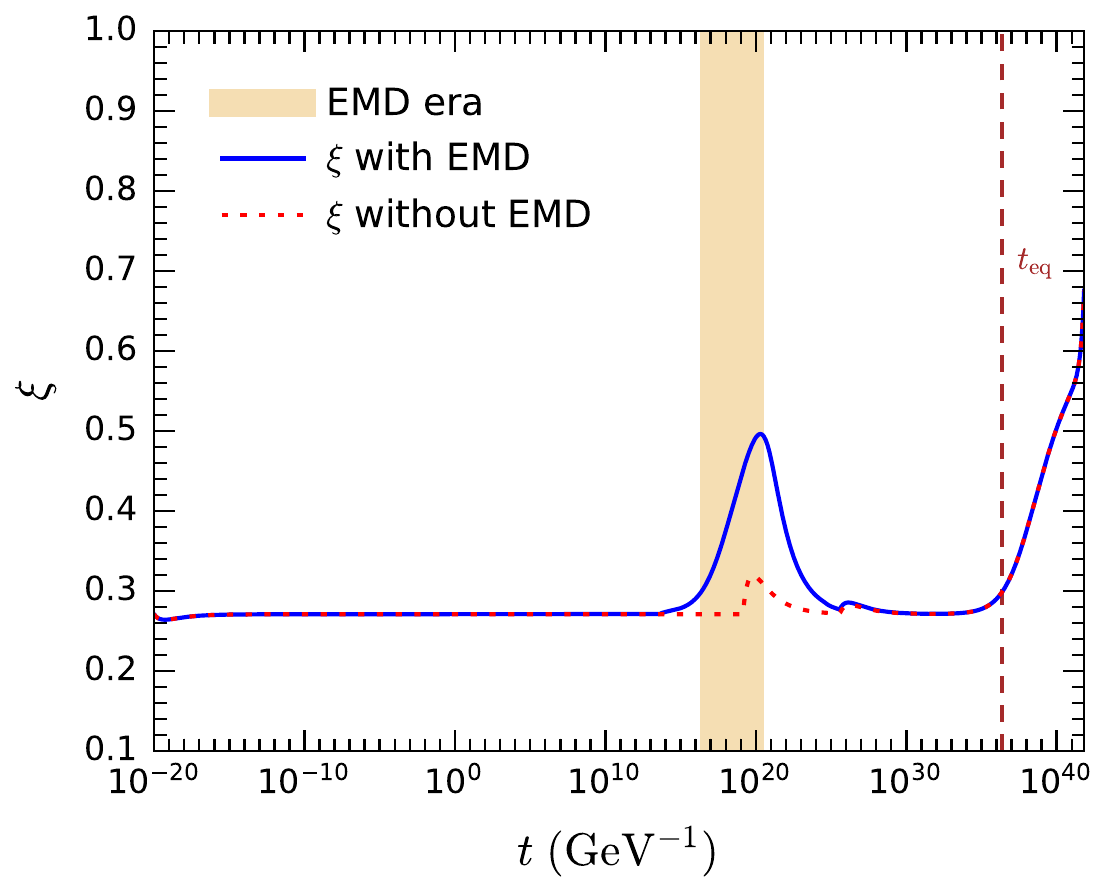}}
	\hspace{.01\textwidth}
	\subfigure[$v(t)$ for BP2a\label{fig:vMH}]{\includegraphics[width=0.48\textwidth]{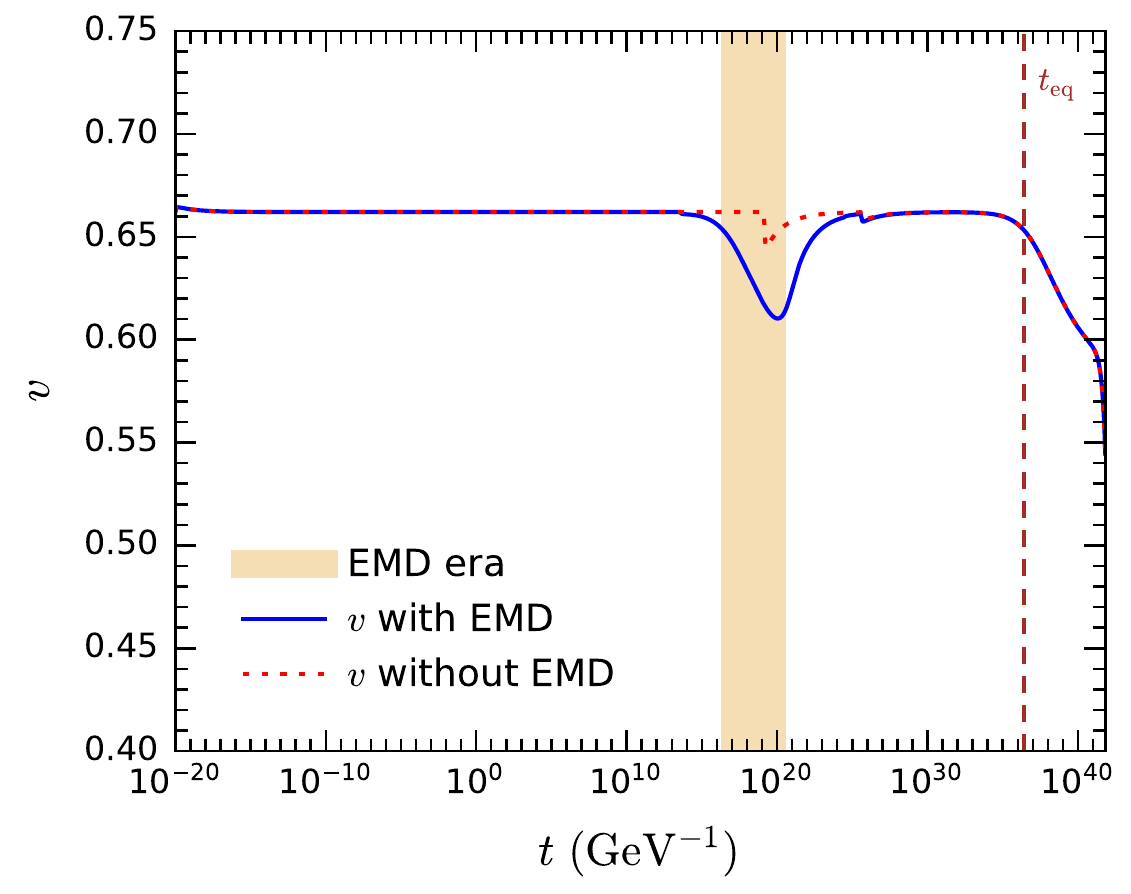}}
    \caption{Evolution of the normalized correlation length $\xi$ (left panels) and the RMS velocity $v$ (right panels) of the CS network in the presence of the EMD era (blue solid lines) for BP1a (upper panels) and BP2a (lower panels). The wheat-colored bands denote the duration of the EMD era, while the vertical dashed lines indicate $t_\mathrm{eq}$. For comparison, the results from the standard $\Lambda\mathrm{CDM}$ model are also plotted as red dotted lines.}
 \label{fig:ksi_v}
\end{figure}

Since the insertion of the EMD era affects the evolution of the CS network in the VOS model, it is necessary to examine how the normalized correlation length $\xi$ and RMS velocity $v$ change with time. We should consider these variations when calculating the number density of CS loops during the EMD era.
To achieve this, we solve Eqs.~\eqref{eqn:ksi} and \eqref{eqn:v} with the Hubble rate $H(t)$ modified by the presence of the EMD era and obtain $\xi$ and $v$ as functions of cosmic time $t$ for BP1a and BP2a, as demonstrated by blue solid lines in Fig.~\ref{fig:ksi_v}.
For comparison, the results of the standard $\Lambda\mathrm{CDM}$ model without the EMD era are presented as red dotted lines.

During the RD era in the $\Lambda\mathrm{CDM}$ model, $\xi$ and $v$ maintain the scaling values $\xi_\mathrm{r} = 0.271$ and $v_\mathrm{r} = 0.662$, as stated in Eq.~\eqref{eqn:scalingregime:RD}.
Some deviations around $t \sim 10^{20}~\si{GeV^{-1}}$ and $t \sim 10^{26}~\si{GeV^{-1}}$ are attributed to changes in relativistic degrees of freedom shown in Eq.~\eqref{eqn:dof}.
For $t \gtrsim t_\mathrm{eq}$, the universe transits to the MD era, resulting in an increase in $\xi$ and a decrease in $v$.
It is important to note that the time axis in Fig.~\ref{fig:ksi_v} is on a logarithmic scale, indicating that the changes in $\xi$ and $v$ occur relatively slowly.
Nonetheless, this represents a mild nonscaling effect.

\begin{figure}[!t]
	\centering
	\subfigure[$t\dot{\xi}/\xi$ for BP1a\label{fig:dksidtHN}]{\includegraphics[width=0.48\textwidth]{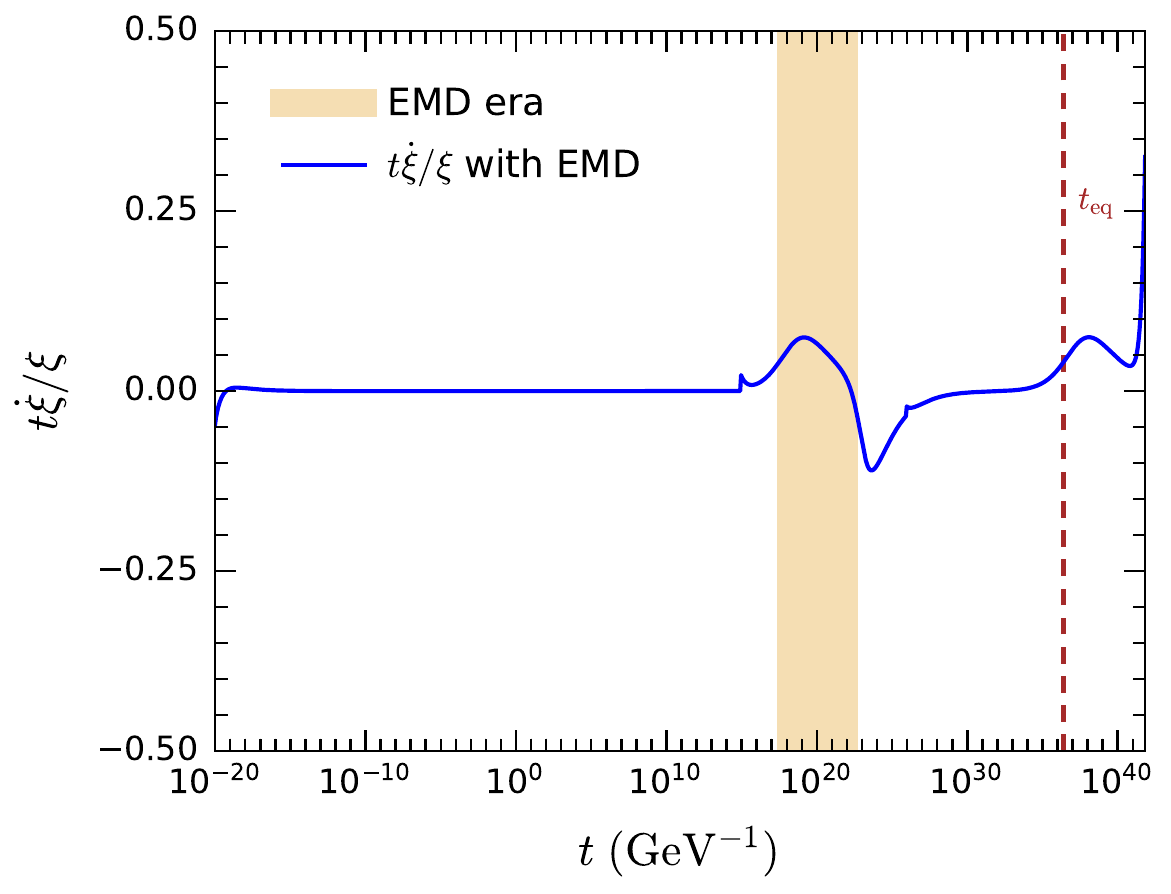}}
	\hspace{.01\textwidth}
	\subfigure[$t\dot{v}/v$ for BP1a\label{fig:dvdtHN}]{\includegraphics[width=0.48\textwidth]{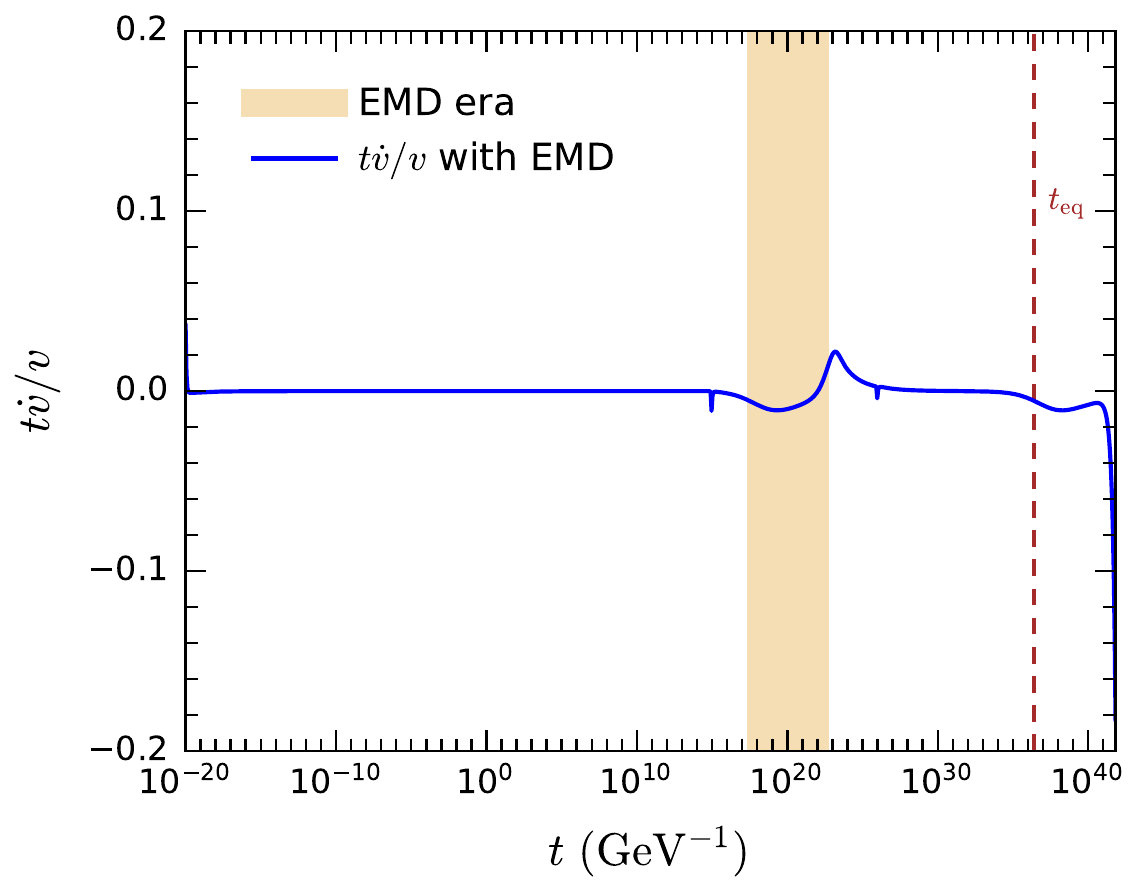}}
    \subfigure[$t\dot{\xi}/\xi$ for BP2a\label{fig:dksidtMH}]{\includegraphics[width=0.48\textwidth]{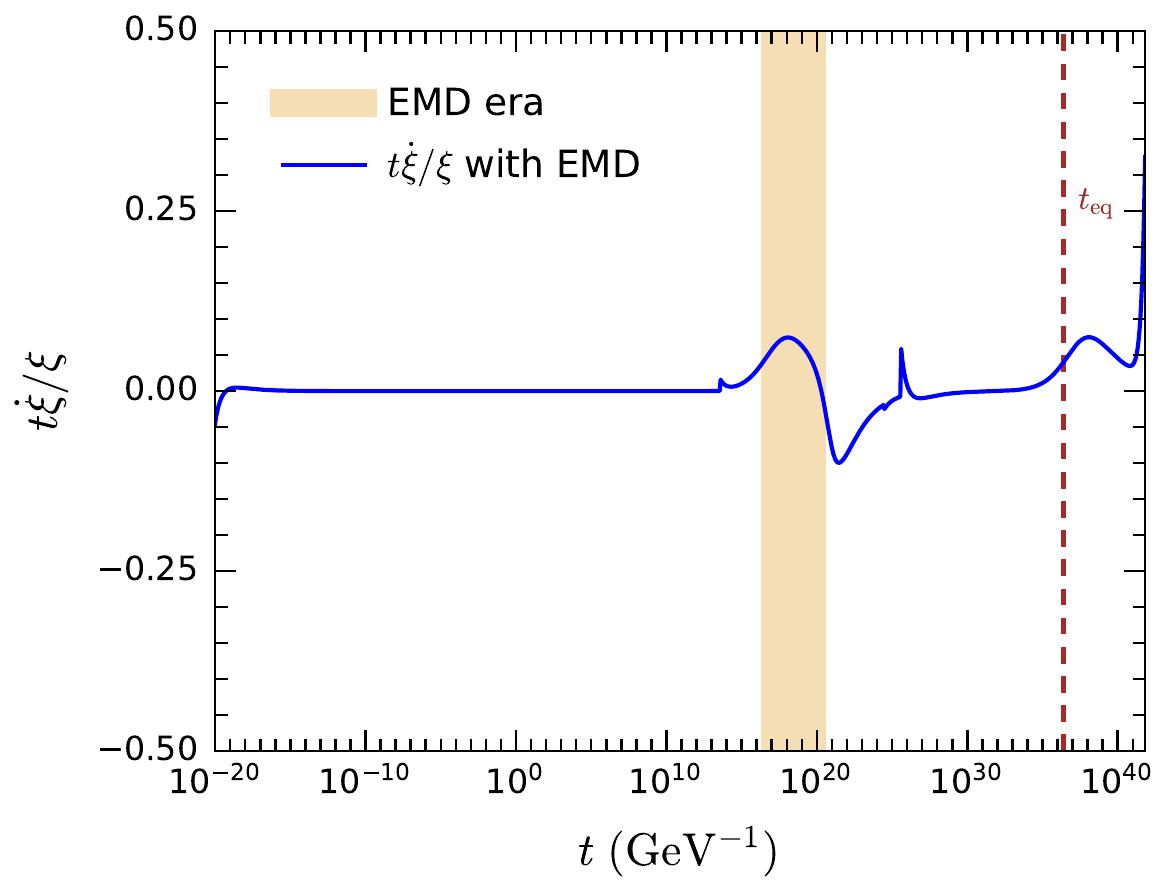}}
	\hspace{.01\textwidth}
	\subfigure[$t\dot{v}/v$ for BP2a\label{fig:dvdtMH}]{\includegraphics[width=0.48\textwidth]{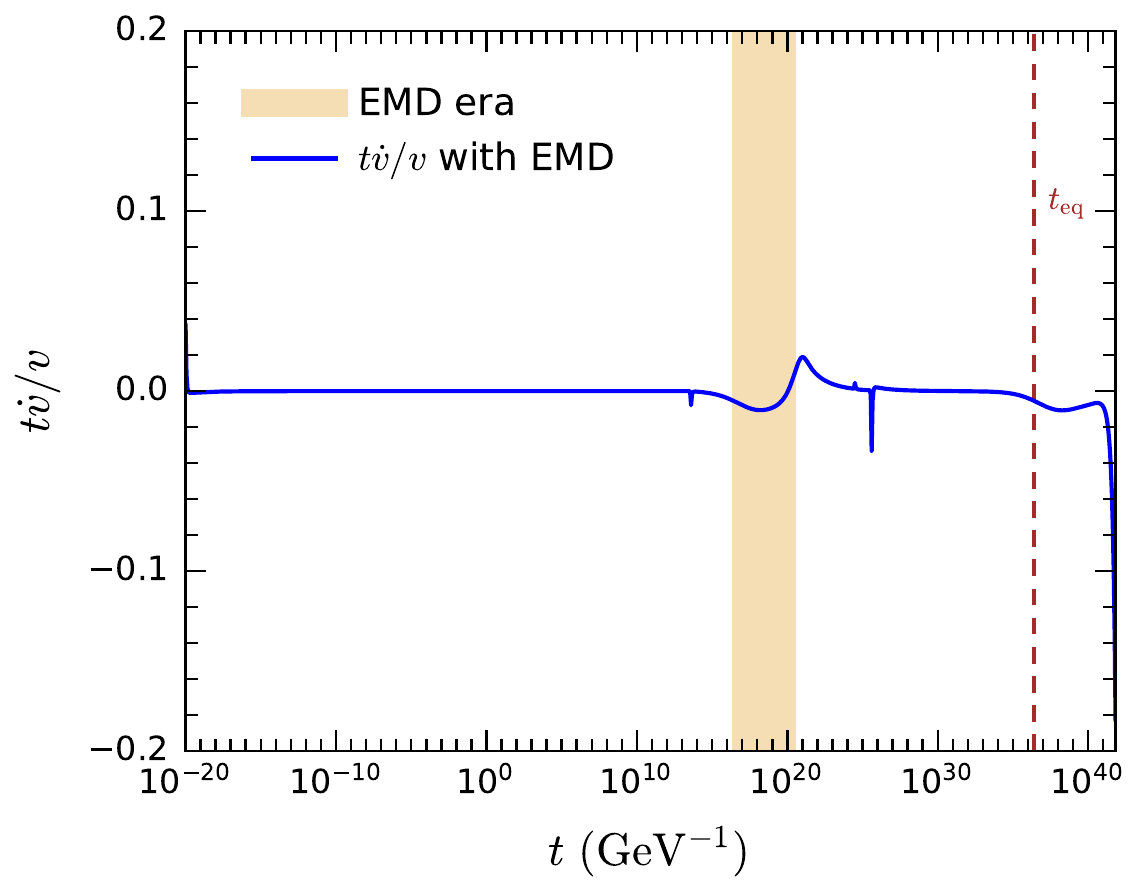}}
    \caption{Evolution of $t\dot{\xi}/\xi$ (left panels) and $t\dot{v}/v$ (right panels) in the presence of the EMD era for BP1a (upper panels) and BP2a (lower panels).}
 \label{fig:dksidt_dvdt}
\end{figure}

When the EMD era occurs, $\xi$ increases, while $v$ decreases, and both quantities gradually return to their scaling values after the EMD era ends.
Therefore, the EMD era also introduces a nonscaling effect, which will be incorporated into the calculation of the CS loop number density.
Furthermore, we estimate the magnitudes of the time derivatives $\dot{\xi}$ and $\dot{v}$ by calculating the evolution of their ratios to $\xi/t$ and $v/t$, respectively, as shown in Fig.~\ref{fig:dksidt_dvdt}.
We find that during the EMD era for both BPs, $|t\dot{\xi}/\xi|$ and $|t\dot{v}/v|$ are smaller than $0.15$ and $0.03$, respectively.
Thus, $\dot{\xi}$ and $\dot{v}$ are negligible compared to $\xi/t$ and $v/t$, simplifying the calculations below.

Following the derivation from Eq.~\eqref{eq:nCSm:1} to Eq.~\eqref{eqn:ncs_ana_m}, we evaluate the CS loop number density formed during the EMD era as
\begin{align}\label{eqn:nEMD}
    n^\mathrm{EMD}_\mathrm{CS}(l,t_1< t\leq t_2) 
    &= \frac{\mathcal{F}_\mathrm{m}\tilde{c}}{t^2}\int_{t_1}^t \frac{v'}{\gamma_{v'}\xi'^3 l'^{1.69}t'^{1.31}}\;\Theta\left(t'-\frac{l'}{\alpha_\mathrm{m}\xi}\right)\mathrm{d}t'\nonumber\\
    &\simeq \frac{\mathcal{F}_\mathrm{m}\tilde{c}}{t^2}\left[-\frac{v'}{0.31\gamma_{v'}\xi'^3l'^{1.69}t'^{0.31}}\bigg|^t_{t_1} + \int_{t_1}^t\frac{v'\delta[t'-l'/(\alpha_\mathrm{m} \xi)]}{0.31\gamma_{v'}\xi'^3l'^{1.69}t'^{0.31}}\mathrm{d}t' \right]\Theta\left(t-\frac{l}{\alpha_\mathrm{m} \xi}\right)\nonumber\\
    &\simeq \frac{\mathcal{F}_\mathrm{m}\tilde{c}}{0.31t^2(l+\Gamma G\mu t)^2} \left[\frac{0.18^{0.31}v_\star}{\gamma_{v_\star}\xi_\star^3} - \frac{(l/t)^{0.31}v}{\gamma_v\xi^3}\right]\Theta(0.18t-l)\nonumber\\
    & ~\quad + \frac{\mathcal{F}_\mathrm{m}\tilde{c} v(t_1)\,\Theta(0.18t_1-l - \Gamma G\mu t)}{0.31\gamma_{v(t_1)}\xi^3(t_1)t_1^2 t^2[l/t_1 + \Gamma G\mu (t/t_1-1)]^{1.69}},
\end{align}
where $\Gamma G \mu \ll \alpha_\mathrm{m}\xi_\star=0.18$ is used at the third step.

In addition, similar to Eq.~\eqref{eqn:nrm}, we must account for the number densities of CS loops produced in a preceding era and surviving into subsequent eras following the timeline in Fig.~\ref{fig:timeaxis}.
For clarity, we define the notations for these loop number densities as follows.
\begin{itemize}
	\item $n_\mathrm{CS}^\mathrm{rm\ast}$: CS loops formed in the original RD era and surviving into the EMD era.
	\item $n_\mathrm{CS}^\mathrm{mr}$: CS loops formed in the EMD era and surviving into the second RD era.
	\item $n_\mathrm{CS}^\mathrm{rm}$: CS loops formed in the second RD era and surviving into the final MD era.
	\item $n_\mathrm{CS}^\mathrm{rmr}$: CS loops formed in the original RD era and surviving into the second RD era.
	\item $n_\mathrm{CS}^\mathrm{mrm}$: CS loops formed in the EMD era and surviving into the final MD era.
	\item $n_\mathrm{CS}^\mathrm{rmrm}$: CS loops formed in the original RD era and surviving into the final MD era.
\end{itemize}

Following the approach in Ref.~\cite{Blanco-Pillado:2013qja} for accounting for the cosmological expansion, $n_\mathrm{CS}^\mathrm{rm}$ can be determined by
\begin{equation}
n_\mathrm{CS}^\mathrm{rm}(l, t> t_\mathrm{eq})a^3(t) = n_\mathrm{CS}^\mathrm{r}(l_\mathrm{eq},t_\mathrm{eq})a^3(t_\mathrm{eq}),
\end{equation}
where $l_\mathrm{eq} = l + \Gamma G\mu (t-t_\mathrm{eq})$.
This leads to
\begin{align}\label{eqn:deduce}
    n_\mathrm{CS}^\mathrm{rm}(l, t> t_\mathrm{eq}) = \frac{t_\mathrm{eq}^2}{t^2}\frac{0.18\,\Theta\{0.1t_\mathrm{eq}-[l+\Gamma G\mu(t-t_\mathrm{eq})]\}}{t_\mathrm{eq}^{{3}/{2}}(l + \Gamma G\mu t)^{{5}/{2}}}\simeq \frac{0.18t_\mathrm{eq}^{{1}/{2}}\Theta(0.1t_\mathrm{eq}-l-\Gamma G\mu t)}{t^2(l + \Gamma G\mu t)^{{5}/{2}}},
\end{align}
which is equivalent to Eq.~\eqref{eqn:nrm}.
Similarly, by introducing $l_{1,2} = l + \Gamma G\mu (t-t_{1,2})$, we obtain
\begin{align}
    n_\mathrm{CS}^\mathrm{rm\ast}(l, t_1 < t \leq t_2) &= \frac{a^3(t_1)}{a^3(t)}\, n_\mathrm{CS}^\mathrm{r}(l_1, t_1)\simeq \frac{0.18t_1^{{1}/{2}}\Theta(0.1t_1-l-\Gamma G\mu t)}{t^2(l + \Gamma G\mu t)^{{5}/{2}}},\\
    n_\mathrm{CS}^\mathrm{rmr}(l, t_2< t\le t_\mathrm{eq}) &=  \frac{a^3(t_2)}{a^3(t)}\, n_\mathrm{CS}^\mathrm{rm\ast}(l_2, t_2) \simeq \frac{0.18t_1^{{1}/{2}}\Theta(0.1t_1-l-\Gamma G\mu t)}{t_2^{{1}/{2}}t^{{3}/{2}}(l+\Gamma G\mu t)^{{5}/{2}}}, \\
    n_\mathrm{CS}^\mathrm{rmrm}(l, t> t_\mathrm{eq}) &=  \frac{a^3(t_\mathrm{eq})}{a^3(t)}\, n_\mathrm{CS}^\mathrm{rmr}(l_\mathrm{eq}, t_\mathrm{eq})\simeq \frac{0.18t_1^{{1}/{2}}t_\mathrm{eq}^{{1}/{2}}\Theta(0.1t_1-l-\Gamma G\mu t)}{t_2^{{1}/{2}}t^2(l+\Gamma G\mu t)^{{5}/{2}}}.
\end{align}

Moreover, the number densities of CS loops produced during the EMD era and surviving into the second RD and final MD eras are given by
\begin{align}
    n_\mathrm{CS}^\mathrm{mr}(l,t_2< t\le t_\mathrm{eq}) &= \frac{a^3(t_2)}{a^3(t)}\,
    n_\mathrm{CS}^\mathrm{EMD}(l_2,t_2)\nonumber\\
    &\simeq \frac{\mathcal{F}_\mathrm{m}\tilde{c}\,\Theta(0.18t_2-l-\Gamma G\mu t)}{0.31t_2^{1/2}t^{3/2}(l+\Gamma G\mu t)^2} \left[\frac{0.18^{0.31}v_\star}{\gamma_{v_\star}\xi_\star^3} - \frac{(l_2/t_2)^{0.31}v(t_2)}{\gamma_{v(t_2)}\xi^3(t_2)}\right]\nonumber\\
    &~\quad + \frac{\mathcal{F}_\mathrm{m}\tilde{c} v(t_1)\Theta(0.18t_1-l - \Gamma G\mu t)}{0.31\gamma_{v(t_1)}\xi^3(t_1)t_1^2 t_2^{1/2} t^{3/2}[l/t_1 + \Gamma G\mu (t/t_1-1)]^{1.69}},
\label{eqn:nmr}\\
    n_\mathrm{CS}^\mathrm{mrm}(l,t>t_\mathrm{eq}) &= \frac{a^3(t_\mathrm{eq})}{a^3(t)}\,n_\mathrm{CS}^\mathrm{mr}(l_\mathrm{eq},t_\mathrm{eq})\nonumber\\
    &\simeq \frac{\mathcal{F}_\mathrm{m}\tilde{c}t_\mathrm{eq}^{1/2}\,\Theta(0.18t_2-l-\Gamma G\mu t)}{0.31t_2^{1/2} t^2(l+\Gamma G\mu t)^2} \left[\frac{0.18^{0.31}v_\star}{\gamma_{v_\star}\xi_\star^3} - \frac{(l_2/t_2)^{0.31}v(t_2)}{\gamma_{v(t_2)}\xi^3(t_2)}\right]\nonumber\\
    &~\quad+ \frac{\mathcal{F}_\mathrm{m}\tilde{c}t_\mathrm{eq}^{1/2} v(t_1)\,\Theta(0.18t_1-l - \Gamma G\mu t)}{0.31\gamma_{v(t_1)}\xi^3(t_1)t_1^2 t_2^{1/2} t^2[l/t_1 + \Gamma G\mu (t/t_1-1)]^{1.69}}.
    \label{eqn:nmrm}
\end{align}

\subsection{Modification of the GW spectrum}

\begin{figure}[!t]
	\centering
	\subfigure[BP1a in Scenario~1\label{fig:CSGWBOSHN}]{\includegraphics[width=0.48\textwidth]{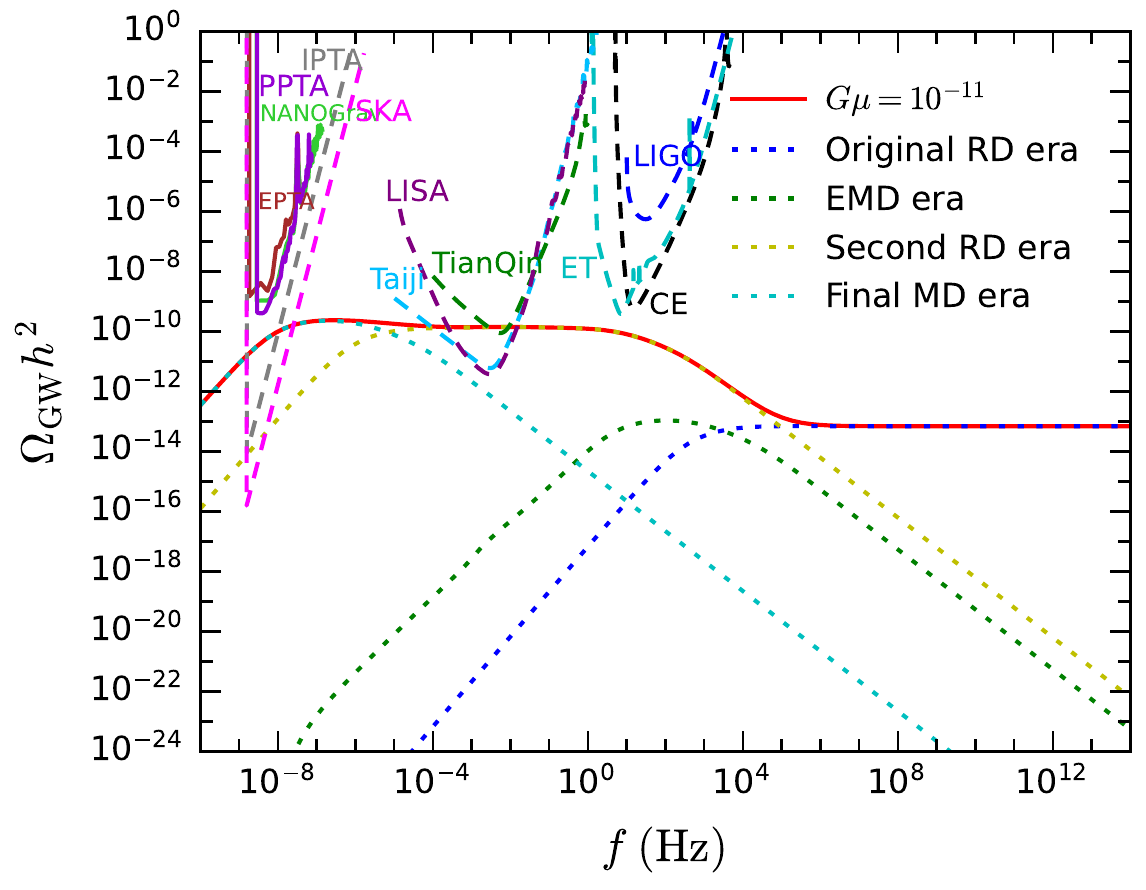}}
	\hspace{.01\textwidth}
	\subfigure[BP2a in Scenario~2\label{fig:CSGWBOSMH}]{\includegraphics[width=0.48\textwidth]{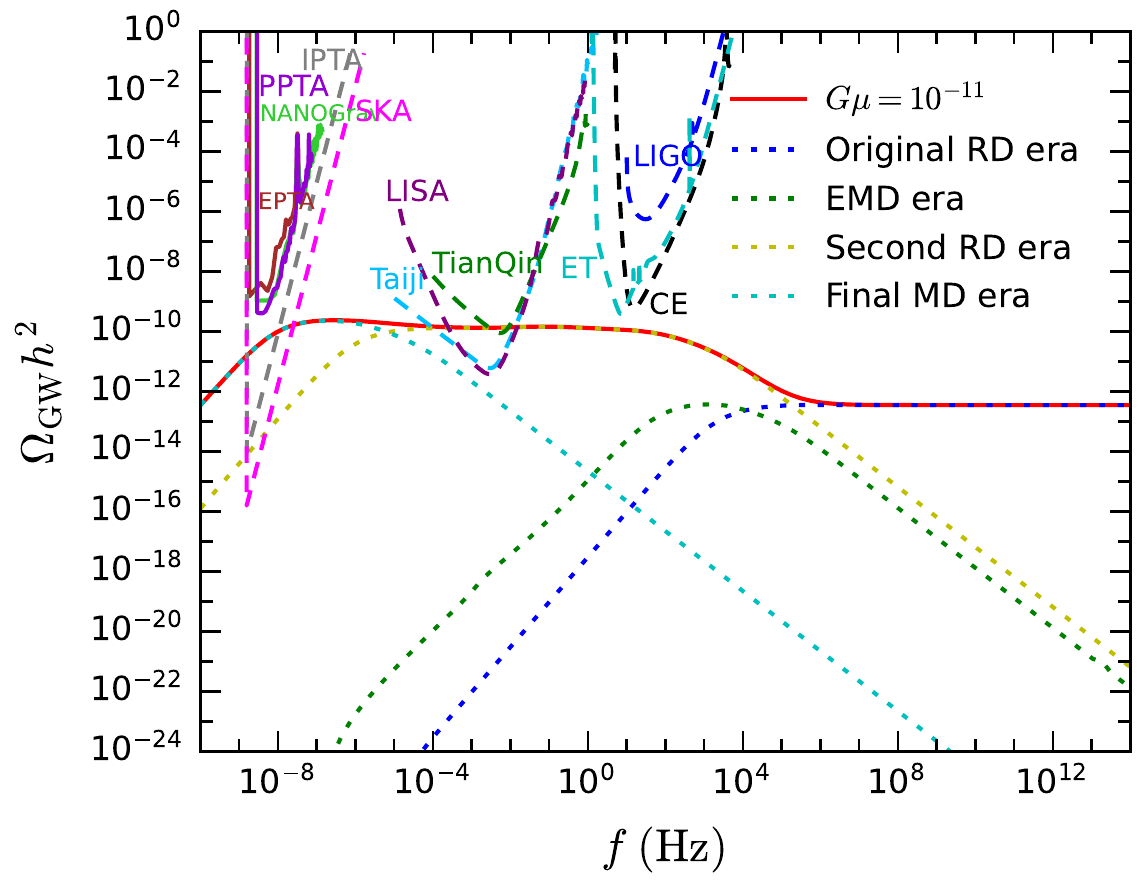}}
	\caption{GW spectra from cosmic strings with $G\mu = 10^{-11}$ modified by the EMD era for BP1a in Scenario 1 (a) and BP2a in Scenario 2 (b). The red solid lines represent the total GW spectra, which include contributions from CS loops across all eras. The blue, green, olive, and cyan dotted lines correspond to the contributions from CS loops formed during the original RD, EMD, second RD, and final MD eras, respectively. For comparison, various constraints and sensitivity curves from several GW detection experiments are also shown.}
 \label{fig:CSGW}
\end{figure}

We now analyze the SGWB spectrum generated by cosmic strings, incorporating modifications to the scale factor $a(t)$ and the CS loop number density $n_\mathrm{CS}(l,t)$ induced by the EMD era.
Based on Eqs.~\eqref{eqn:Cn} and \eqref{eqn:OGW}, the GW spectra for $G\mu = 10^{-11}$ are presented in Figs.~\ref{fig:CSGWBOSHN} and \ref{fig:CSGWBOSMH} for BP1a and BP2a, respectively.
The contributions from CS loops formed during different eras are demonstrated separately.
Note that, in earlier eras, the CS loops are smaller in length and generate GWs at higher frequencies.

To assess the sensitivity of GW detection experiments, we also show the constraints from the North American Nanohertz Observatory for Gravitational Waves (NANOGrav)~\cite{NANOGRAV:2018hou}, European Pulsar Timing Array (EPTA)~\cite{Lentati:2015qwp}, and Parkes Pulsar Timing Array (PPTA)~\cite{Shannon:2015ect}, as well as the sensitivity curves of the International Pulsar Timing Array (IPTA)~\cite{Hobbs:2009yy}, Square Kilometer Array (SKA)~\cite{Janssen:2014dka}, Laser Interferometer Space Antenna (LISA)~\cite{LISA:2017pwj}, TianQin~\cite{Liang:2021bde}, Taiji~\cite{Ruan:2018tsw}, LIGO~\cite{KAGRA:2013rdx}, Cosmic Explorer (CE)~\cite{LIGOScientific:2016wof}, and Einstein Telescope (ET)~\cite{Hild:2010id}.

Compared to the GW spectrum for $G\mu = 10^{-11}$ in the $\Lambda\mathrm{CDM}$ model illustrated in Fig.~\ref{fig:CSGW_ori}, the spectra in both BPs with the EMD era display a suppression at high frequencies $f \gtrsim 10~\si{Hz}$, which corresponds to the contributions from CS loops formed during the original RD and EMD eras.
The primary reason for this suppression is as follows.
Since the scaling behavior of the CS network is only slightly violated, the correlation length $L$ remains approximately proportional to the cosmic time $t$, suggesting that the lengths of the generated loops are positively correlated with the scale factor $a(t)$.
As shown in Fig.~\ref{fig:a_h_t}, the EMD era reduces the scale factor before $t_2$.
This means that CS loops with a given initial length $l$, which is inversely proportional to the frequencies of the emitted GWs through $f_\mathrm{e} = 2n/l$, are formed at a later time when the energy densities of both CS loops and emitted GWs are reduced.
Consequently, the GW spectrum at sufficiently high frequencies is suppressed, and the end time of the EMD era, $t_2$, is related to the GW frequency at which this suppression effect becomes significant.
For the parameters listed in Table~\ref{tab:para1}, the dilutor $N_2$ in BP1a has a longer lifetime than the dilutor $\Delta$ in BP2a, resulting in a longer EMD era and, consequently, a stronger suppression.

\begin{figure}[!t]
	\centering
	\subfigure[BP1a in Scenario~1\label{fig:CSGWHN2}]{\includegraphics[width=0.48\textwidth]{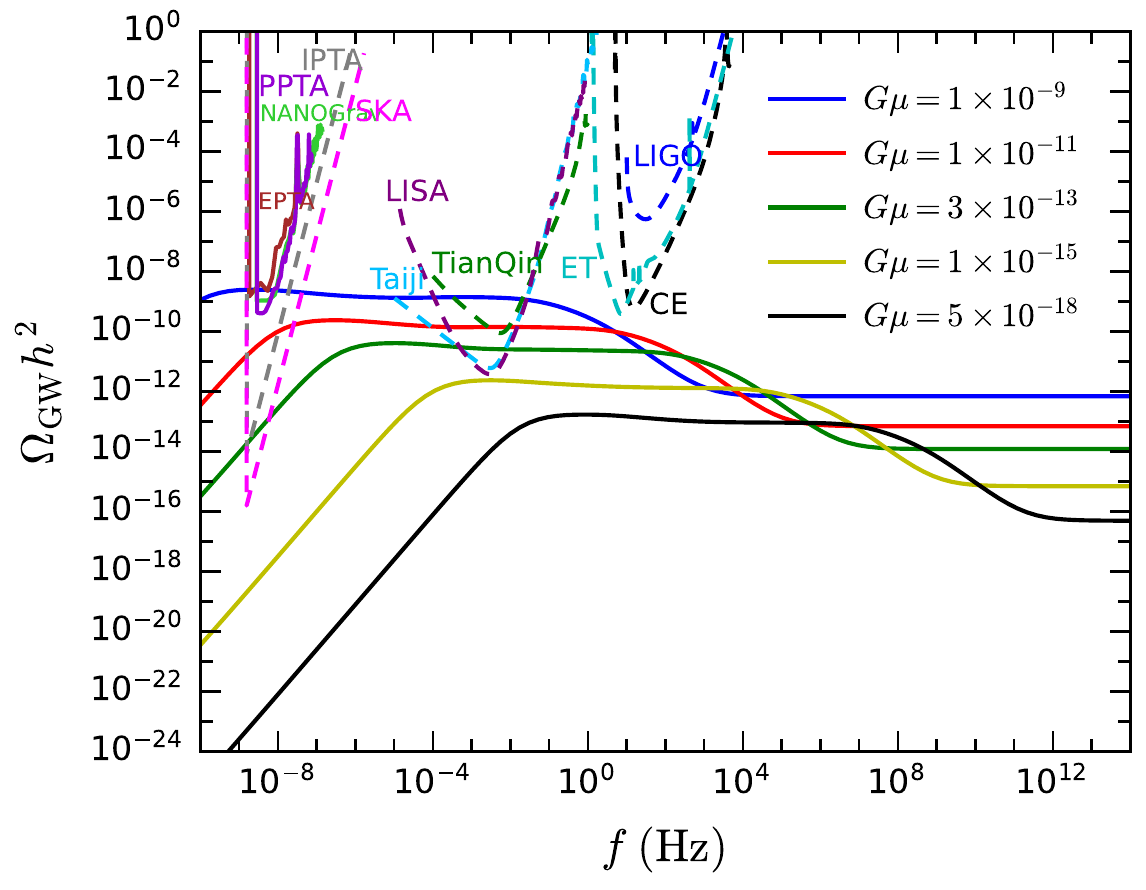}}
	\hspace{.01\textwidth}
	\subfigure[BP2a in Scenario~2\label{fig:CSGWMH2}]{\includegraphics[width=0.48\textwidth]{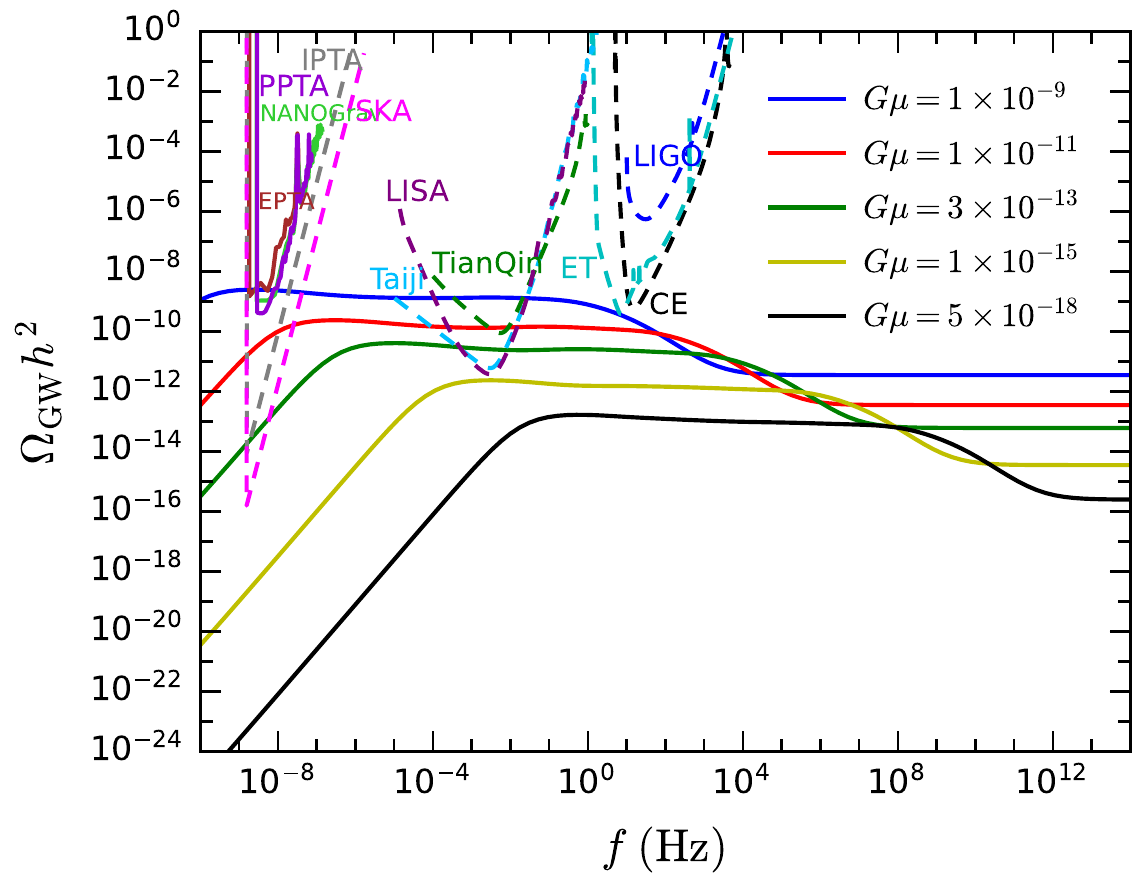}}
	\caption{GW spectra modified by the EMD era for BP1a in Scenario~1 (a) and BP2a in Scenario~2 (b) assuming different values of the CS tension parameter $G\mu$.}
 \label{fig:CSGW2}
\end{figure}

We proceed to investigate how the GW spectrum changes with different parameters.
In Fig.~\ref{fig:CSGW2}, the GW spectra for BP1a and BP2a are displayed with the CS tension parameter varying as $G\mu = 10^{-9}$, $10^{-11}$, $3\times 10^{-13}$, $10^{-15}$, and $5\times 10^{-18}$.
For a smaller CS tension, the GW emission power is reduced, and the lifetimes of CS loops are extended.
This suggests that the loops existing at $t_2$ originated from earlier times with smaller initial lengths, reducing the average loop length at $t_2$ and increasing the frequencies of the emitted GWs.
Consequently, the suppression effect caused by the EMD era begins at a higher GW frequency, as shown in Fig.~\ref{fig:CSGW2}.

\begin{figure}[!t]
	\centering
	\subfigure[Scenario 1\label{fig:CSGWcompHN}]{\includegraphics[width=0.48\textwidth]{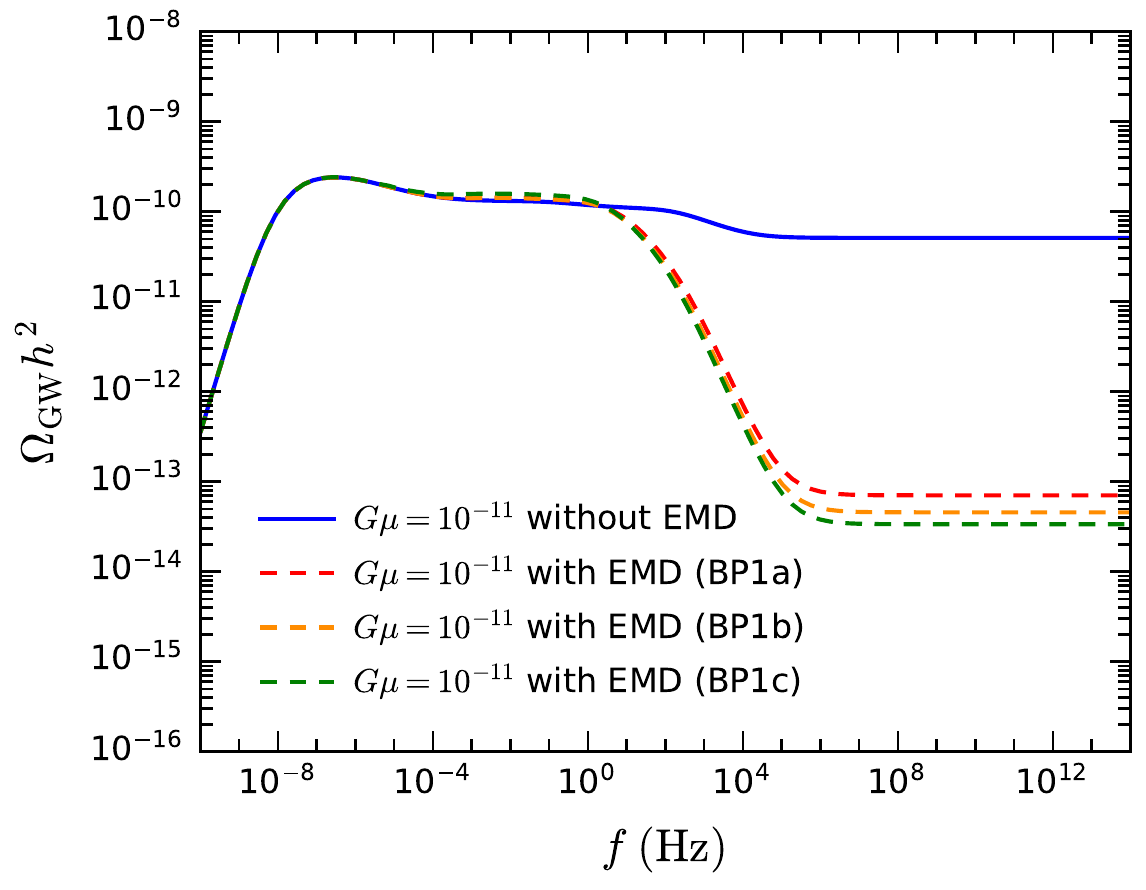}}
	\hspace{.01\textwidth}
	\subfigure[Scenario 2\label{fig:CSGWcompMH}]{\includegraphics[width=0.48\textwidth]{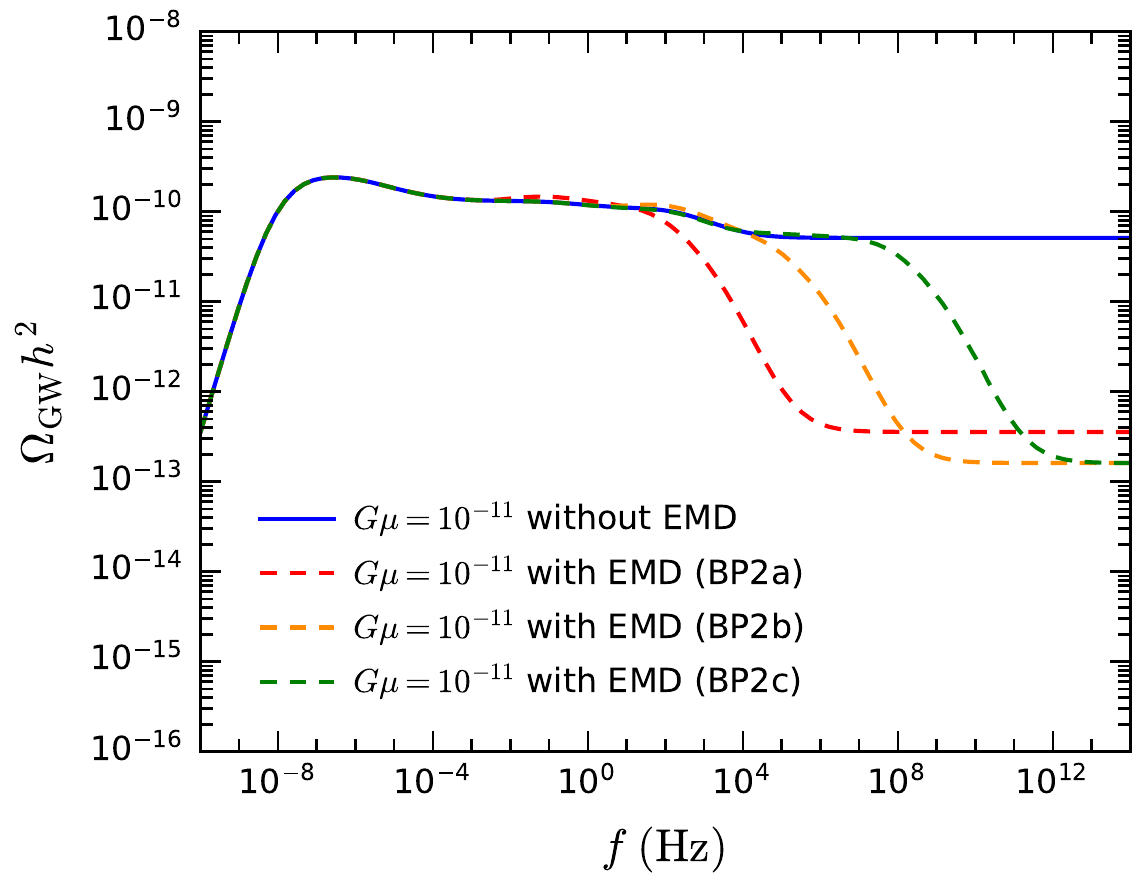}}
	\caption{GW spectra with $G\mu=1\times 10^{-11}$ modified by the EMD era for various BPs in Scenarios 1 (a) and 2 (b). The blue solid lines represents the corresponding GW spectrum in the standard $\Lambda\mathrm{CDM}$ model without the EMD era.}
    \label{fig:CSGWcomparison}
\end{figure}

In addition, we consider the results for BP1b and BP1c in Scenario~1 and for BP2b and BP2c in Scenario~2, whose parameters are tabulated in Tables~\ref{tab:para1} and \ref{tab:para2}, respectively.
The obtained GW spectra are illustrated in Fig.~\ref{fig:CSGWcomparison}.
For BP1a, BP1b, and BP1c in Scenario~1, the dilutor $N_2$ has the same mass but the masses of the DM candidate $N_1$ differ, causing variations in the decay width of $N_2$.
A smaller decay width $\Gamma_{N_2}$ corresponds to a longer duration of the EMD era, leading to stronger suppression effects at high frequencies, as illustrated in Fig.~\ref{fig:CSGWcompHN}.
In contrast, the masses of the dilutor $\Delta$ are different for BP2a, BP2b, and BP2c in Scenario~2.
A heavier $\Delta$ implies that the EMD era occurs earlier, and hence, a higher frequency at which the GW spectrum begins to be suppressed.
This is clearly demonstrated in Fig.~\ref{fig:CSGWcompMH}.

Furthermore, we assess the influence of the suppression effect caused by the EMD era on the experimental sensitivity of ground-based interferometers.
Given a GW spectrum $\Omega_\mathrm{noise}(f)$ converted from the strain noise of a interferometer system, along with a signal spectrum $\Omega_\mathrm{signal}(f)$, the signal-to-noise ratio (SNR) for a practical observational time $t_\mathrm{obs}$ can be evaluated as~\cite{Thrane:2013oya, Schmitz:2020syl}
\begin{equation}\label{eqn:SNR}
    \varrho = \left\{t_\mathrm{obs}\int_{f_\mathrm{min}}^{f_\mathrm{max}}\left[\frac{\Omega_\mathrm{signal}(f)}{\Omega_\mathrm{noise}(f)}\right]^2\mathrm{d}f\right\}^{1/2},
\end{equation}
where $[f_\mathrm{min}, f_\mathrm{max}]$ is the accessible frequency band.
If the SNR reaches a threshold $\varrho_\mathrm{thr} = 10$, it is probable that a GW signal will be detected.

\begin{figure}[!t]
	\centering
	\subfigure[BP1a in Scenario~1\label{fig:SNRHN}]{\includegraphics[width=0.48\textwidth]{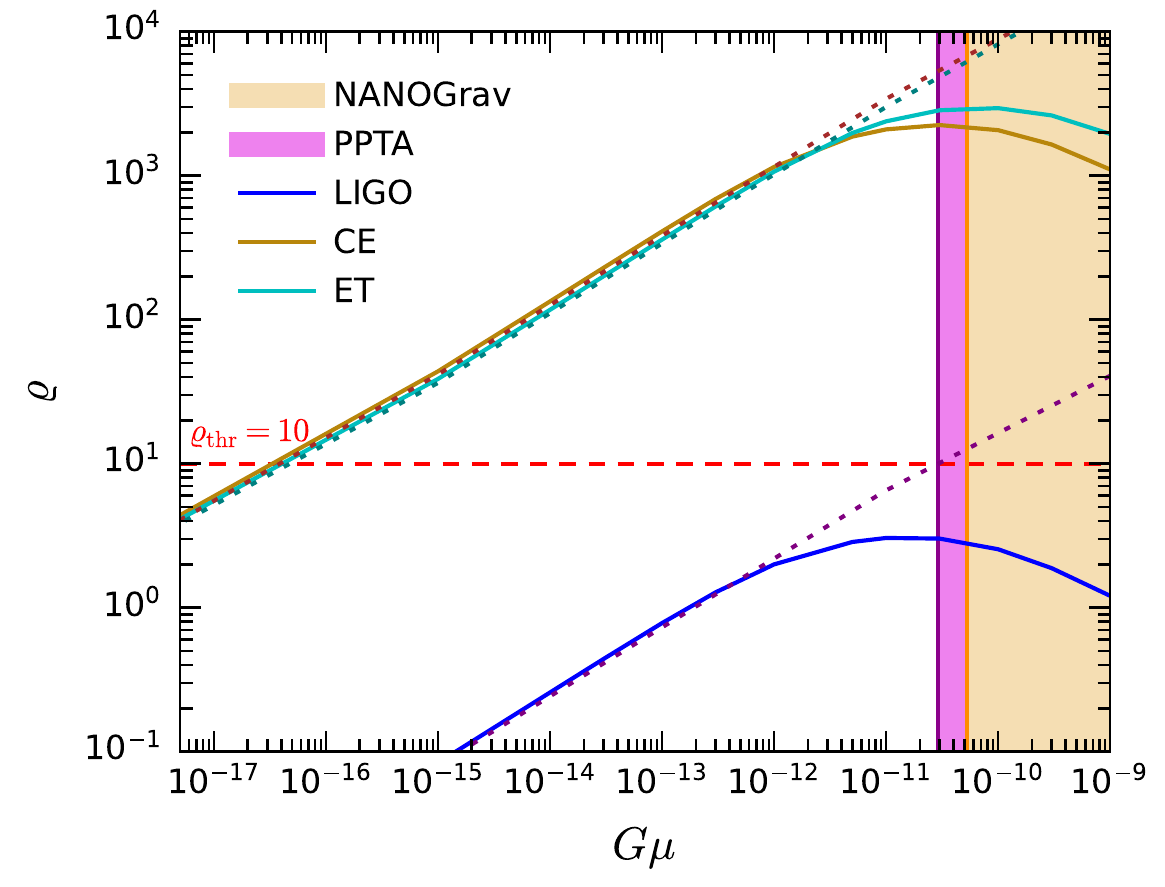}}
	\hspace{.01\textwidth}
	\subfigure[BP2a in Scenario~2\label{fig:SNRMH}]{\includegraphics[width=0.48\textwidth]{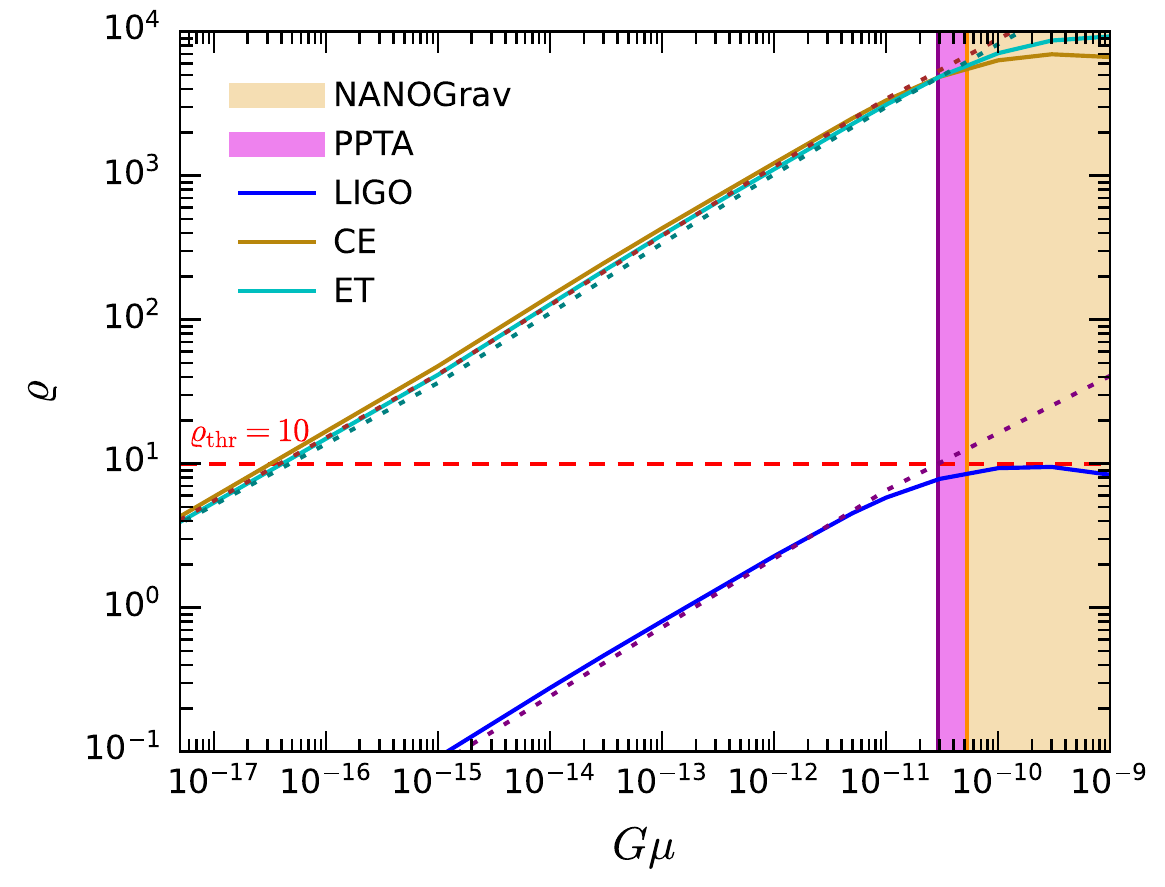}}
	\caption{Estimated SNRs of the LIGO (blue solid lines), CE (gold solid lines), and ET (cyan solid lines) experiments affected by the EMD era for BP1a in Scenario~1 (a) and BP2a in Scenario~2 (b). For comparison, the corresponding SNRs for the $\Lambda\mathrm{CDM}$ models are represented by dotted lines in similar colors.}
 \label{fig:SNR_EMD}
\end{figure}

Using the sensitivity curves of LIGO, CE, and ET, we compute the SNRs for $t_\mathrm{obs}=1~\si{yr}$ as functions of the CS tension parameter $G\mu$.
The results with the EMD era for BP1a in Scenario~1 and BP2a in Scenario~2 are presented by solid lines in Figs.~\ref{fig:SNRHN} and \ref{fig:SNRMH}, respectively, while those for the $\Lambda\mathrm{CDM}$ model are indicated by dotted lines.
In addition, the upper limits $G\mu < 5.3\times 10^{-11}$ and $G\mu < 2.88\times 10^{-11}$ at 95$\%$ confidence level given by NANOGrav~\cite{NANOGRAV:2018hou} and PPTA~\cite{Bian:2022tju} searches for a SGWB generated by cosmic strings are indicated by the regions shaded in wheat and magenta colors.
For the employed BPs, the EMD era does not influence the GW spectrum in the $\sim \si{nHz}$ frequency band, where the PTA experiments are sensitive; thus, these upper limits remain valid.

For $G\mu \lesssim 10^{-12}$, the inclusion of the EMD era has an insignificant impact on the SNRs because the suppression effect occurs at frequencies outside the sensitive bands of LIGO, CE, and ET (cf. Fig.~\ref{fig:CSGW2}).
In contrast, for $G\mu \gtrsim 10^{-12}$, the SNRs are obviously reduced.
Compared to BP2a, the dilutor in BP1a has a longer lifetime, leading to a stronger suppression effect and, consequently, smaller SNRs.

\section{Summary}
\label{sec:sum}

In this study, we explore the influence of an EMD era on the dynamics of a preexisting CS network and the resulting SGWB.
We first review the GWs generated by CS loops, highlighting the explicit dependence of the SGWB spectrum on the CS loop number density $n_\mathrm{CS}$ and the scale factor $a(t)$.
We then analyze the evolution of the CS network based on the VOS model and discuss the scaling behavior.
By assuming appropriate forms for the loop production functions, we derive expressions for the CS loop number densities in RD and MD eras, which are consistent with numerical simulations in the scaling regime.
Notably, these expressions are expected to remain valid even in cases where the scaling behavior is violated.

Next, we consider a cosmic timeline that incorporates an EMD era within the conventional RD era.
As an illustrative example, this EMD era is attributed to a massive, long-lived dilutor in the DM dilution mechanism, particularly within the framework of the minimal LRSM, where the dilutor is either $N_2$ in Scenario~1 or $\Delta$ in Scenario~2.
In this context, the overproduction of the DM particle $N_1$ is resolved by entropy injection from the dilutor decays.
By solving the Boltzmann equations, we obtain the evolution of the number densities for the dilutor, DM, and SM particles, as well as the time dependence of the scale factor.
These solutions allow us to identify the onset and conclusion of the EMD era.
Compared to the standard $\Lambda\mathrm{CDM}$ cosmological model, the scale factor before the end time $t_2$ of the EMD era is smaller.

Furthermore, we show that the evolution of the normalized correlation length $\xi$ and the RMS velocity $v$ of the CS network is modified by the presence of the EMD era, exhibiting nonscaling behavior.
Subsequently, we calculate the number density of CS loops formed during the EMD era, as well as the number densities of loops generated in earlier eras and surviving into later eras.
Using these results, the influence of the EMD era on the SGWB spectrum arising from CS loops is demonstrated.

We find that the EMD era induces a suppression in the SGWB spectrum at sufficiently high frequencies, which is linked to the reduction of the scale factor prior to $t_2$.
Additionally, for a smaller CS tension, the average length of CS loops at $t_2$ is smaller, causing the suppression effect to begin at a higher frequency.
Moreover, a smaller decay width of the dilutor implies a longer duration of the EMD era, leading to a stronger suppression effect.
Furthermore, a heavier dilutor causes an EMD era to occur earlier, resulting in suppression at higher frequencies.

Finally, we estimate the SNRs of the ground-based interferometers LIGO, CE, and ET as functions of $G\mu$.
The results show that for BP1a and BP1b with $G\mu \gtrsim 10^{-12}$, the SNRs are significantly reduced, as the suppression effect falls within the sensitive frequency bands of these experiments.
This study highlights that changes in the cosmic history, such as the presence of an EMD era, can affect the dynamics of the CS network and produce observable signatures in the resulting SGWB spectrum.

\begin{acknowledgments}

The authors acknowledge Ye-Ling Zhou for helpful discussions.
This work is supported by the Guangzhou Science and Technology Planning Project under Grant No.~2024A04J4026.

\end{acknowledgments}

\bibliographystyle{utphys}
\bibliography{ref}
\end{document}